\documentclass[12pt]{iopart}
\usepackage{iopams} 
\usepackage{graphicx} 
\usepackage[hyphens]{url}
\usepackage{cite}

\newcommand{\ket}[1]{\left|#1\right\rangle}

\begin{document}

\title[Advances in the PTB primary fountain clocks]{Advances in the accuracy, stability, and reliability of the PTB primary fountain clocks}

\author{S Weyers$^1$, V Gerginov$^1$\footnote{Present address:  Time and Frequency Division, NIST, 325 Broadway, Boulder, CO 80305, USA
}, M Kazda$^1$, J Rahm$^1$, B Lipphardt$^1$, G Dobrev$^1$\footnote{Present address:  Faculty of Physics, Sofia University ``St. Kliment Ohridski'', 5 J. Bourchier blvd., 1164 Sofia, Bulgaria} and K Gibble$^2$}

\address{$^1$Physikalisch-Technische Bundesanstalt (PTB), Bundesallee 100, 38116 Braunschweig, Germany} 
\address{$^2$Department of Physics, The Pennsylvania State University, University Park, PA 16802, USA} 

\ead{stefan.weyers@ptb.de}
\vspace{10pt}
\begin{indented}
\item
\end{indented}

\begin{abstract}

Improvements of the systematic uncertainty, frequency instability, and long-term reliability of the two caesium fountain primary frequency standards CSF1 and CSF2 at PTB (Physikalisch-Technische Bundesanstalt) are described. We have further investigated many of the systematic effects and made a number of modifications of the fountains. With an optically stabilized microwave oscillator, the quantum projection noise limited frequency instabilities are improved to $7.2 \times 10^{-14} (\tau/1\,\mathrm{s})^{-1/2}$ for CSF1 and $2.5 \times 10^{-14} (\tau/1\,\mathrm{s})^{-1/2}$ for CSF2 at high atom density. The systematic uncertainties of CSF1 and CSF2 are reduced to $2.74 \times 10^{-16}$ and $1.71 \times 10^{-16}$, respectively. Both fountain clocks regularly calibrate the scale unit of International Atomic Time (TAI) and the local realization of Coordinated Universal Time, UTC(PTB), and serve as references to measure the frequencies of local and remote optical frequency standards.

\end{abstract}

\vspace{2pc}

\maketitle

\section{Introduction}

For nearly two decades microwave caesium fountain frequency standards \cite{Wynands2005} realize most accurately the SI-second. Currently, the scale unit of International Atomic Time (TAI) is almost exclusively based on monthly frequency calibration reports by several fountain clocks serving as primary frequency standards (PFS) \cite{CircT}. During the past five years, eleven primary caesium fountain clocks from eight different metrology institutes in Europe \cite{Weyers2001,Gerginov2010,Szymaniec2010,Guena2012,Domnin2013, Levi2014}, the U.S. \cite{Jefferts2002,Heavner2014,Gibble2015} and Asia \cite{Fang2015,Acharya2017} contributed to TAI. One of these fountains additionally operated as a dual fountain, simultaneously using caesium and rubidium atoms \cite{Guena2014}, providing additional TAI calibrations with a secondary representation of the SI second.

We have investigated several systematic effects and have made a number of improvements to the fountain clocks CSF1 and CSF2 at the Physikalisch-Technische Bundesanstalt (PTB). 
As a result of recent modifications and further investigations, both fountains have reached a degree of maturity and are approaching anticipated ultimate performance levels for their frequency instabilities, systematic uncertainties and long-term reliabilities. This has been evinced by the quality and number of TAI calibrations and optical frequency measurements \cite{Huntemann2014,Tamm2014,Grebing2016,Matveev2013,Friebe2011}. In addition, the local realization of Coordinated Universal Time (UTC), UTC(PTB), is maintained for a number of years by a hydrogen maser that is steered daily by fountain measurements \cite{Bauch2012}. Recently, the first comparison of distant fountain clocks, at PTB and LNE-SYRTE (Laboratoire National de m\'{e}trologie et d'Essais - SYst\`{e}me de R\'{e}f\'{e}rences Temps-Espace) in Paris, via a 1400~km long optical fibre link showed very good agreement, below the $3 \times 10^{-16}$ level for all of the participating fountain clocks, which is compatible with their statistical and systematic uncertainties \cite{Guena2017}.

Here, we update the status and the full accuracy evaluations of CSF1 and CSF2. We begin with a description of the main features and recent modifications of the fountains (section \ref{sec:description}). We then give a comprehensive update of the systematic uncertainties (section \ref{sec:Systematics}) and subsequently discuss the fountain frequency instabilities, in conjunction with the achievable statistical uncertainties (section \ref{sec:Instabil}). Finally, we compile the recent applications of CSF1 and CSF2, as well as corresponding measurements (section \ref{sec:Applications}).

\section{Features and modifications of the CSF1 and CSF2 fountain clocks}
\label{sec:description}

\begin{figure}[t]
  \includegraphics[width=\columnwidth]{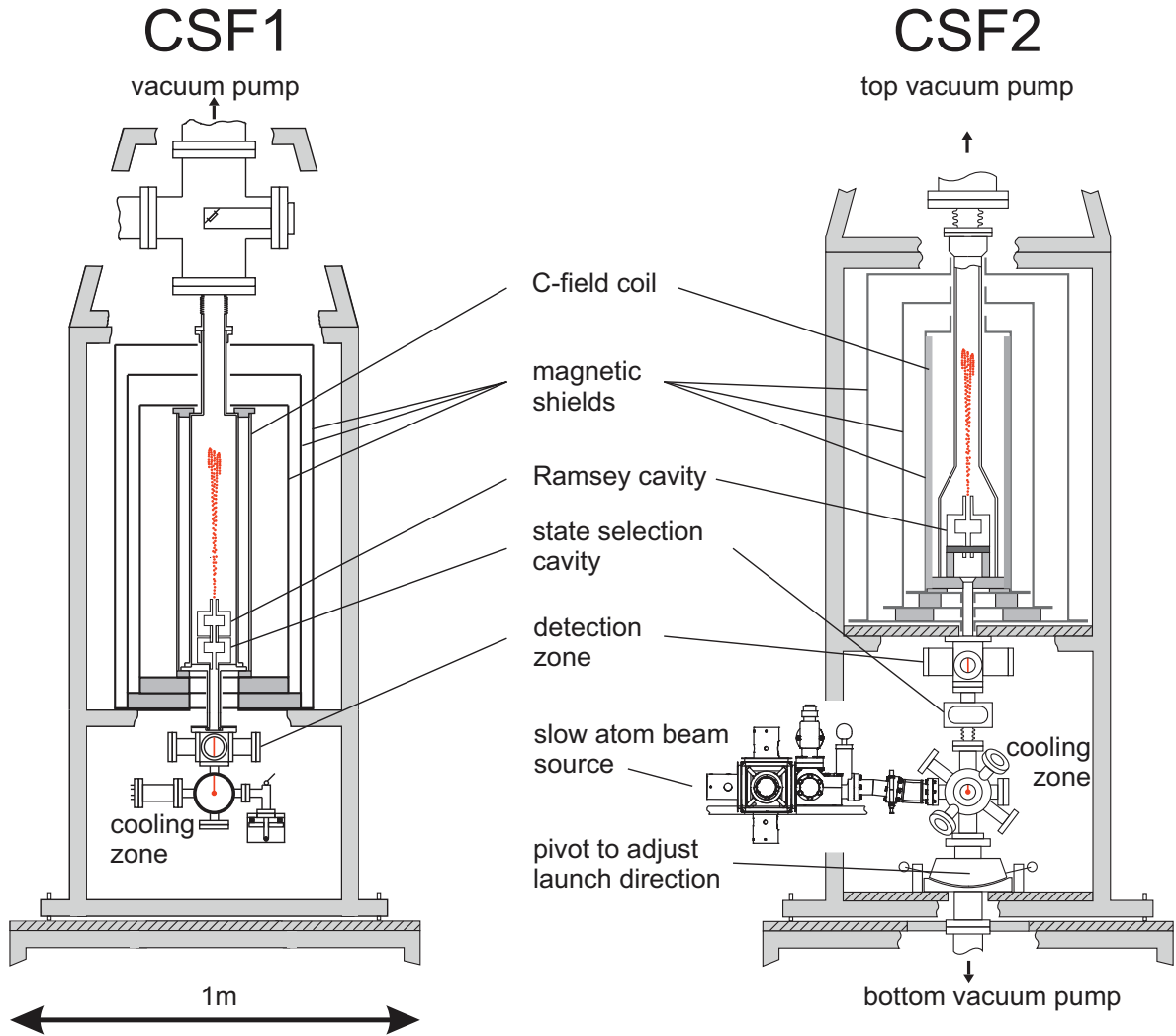}
	\caption{Schematics of the caesium fountain clocks CSF1 and CSF2 of PTB.}
	\label{fig:CSF1CSF2} 
\end{figure}

Figure~\ref{fig:CSF1CSF2} is a schematic of the physics packages of both fountains. The first detailed descriptions appeared in \cite{Weyers2001,Weyers2001b} for CSF1 and in \cite{Gerginov2010} for CSF2. Next, we summarize significant features and modifications specific to each fountain, and then report modifications common to both.

\subsection{CSF1}

CSF1 first operated as a PFS in August 2000. Since then, more than 70 TAI calibration reports were submitted (see subsection~\ref{sec:TAI}), and since 2010, CSF1 steers UTC(PTB). 

In CSF1 we collect atoms in a magneto-optical trap (MOT) from the background Cs vapour, for typically $\approx 160$\,ms. This loading time is a reasonable compromise between collecting many atoms to lower the frequency instability (see section~\ref{sec:Instabil}) without a large collisional shift uncertainty (see subsection~\ref{subsec:Coll}). Its precise value is chosen to average out potential microwave phase deviations synchronous with the fountain cycle (see subsection~\ref{subsec:Electr}). Following the MOT loading and an optical molasses phase, the atoms are accelerated to 4.04\,m/s within 1\,ms by laser detuning in the (0,0,1) geometry, as in figure~\ref{fig:CSF1CSF2}, and cooled to $1.8\,\mu$K within 1.3\,ms. After 93\,ms the atoms reach the rectangular TE$_{201}$ state selection cavity in the magnetically shielded interaction region, where they are prepared in the $\ket{F=3, m_F=0}$ state \cite{Wynands2005,Weyers2001b}. They reach the cylindrical TE$_{011}$ Ramsey cavity \cite{Schroeder2002}, mounted just above the state selection cavity, at 125\,ms after launch. The atoms reach a maximum height of 0.83\,m, with $T_\mathrm{R}=564$\,ms between the first passage of the Ramsey cavity and the second downward passage, which gives an 0.89\,Hz fullwidth at half maximum (FWHM) central Ramsey fringe. Finally, the clock state populations are detected, 11\,cm above the cooling zone.

The CSF1 laser system \cite{Weyers2001} uses a homemade extended-cavity diode laser (ECDL) for detection and as the master laser for a single slave laser, which provides the cooling light. The former repumping ECDL was replaced by a distributed Bragg reflector (DBR) laser diode. All laser light is delivered to the fountain by polarization-maintaining optical fibres. In contrast to the former system, the master laser detection light is now delivered to the fountain by a single optical fibre, and then split into a lower and an upper laser beam to detect separately atoms in the $\ket{F=3, m_F=0}$ and $\ket{F=4, m_F=0}$ states. Another fibre delivers the repumping light to the detection zone. As a result, polarization variations at the master laser fibre output produce small common mode power fluctuations of both detection beams, which do not degrade the signal-to-noise. The design of our previous homemade shutters required sensitive alignments to avoid scattered laser light from reaching the optical fibre inputs. After replacing them with commercial shutters, we detect no scattered light at the fibre inputs, as described in subsection~\ref{subsec:ACStark}.

\subsection{CSF2}

CSF2 started operating as a PFS in December 2008.  Since then, more than 60 TAI calibration reports for it were submitted (see subsection~\ref{sec:TAI}). In 2013 CSF2 joined CSF1 to steer UTC(PTB).

A low-velocity intense source (LVIS) in CSF2 provides a cold atom beam to load atoms in optical molasses (figure~\ref{fig:CSF1CSF2}) \cite{Lu1996,Dobrev2016}. The molasses laser beams, with a (1,1,1) geometry, collect atoms for typically 340\,ms, again, as a compromise between the frequency instability (see section~\ref{sec:Instabil}) and the collisional shift uncertainty (see subsection~\ref{subsec:Coll}), and to average out potential microwave phase deviations synchronous with the fountain cycle (see subsection~\ref{subsec:Electr}). After molasses loading, the LVIS laser beams are blocked and the atoms are accelerated to 4.43\,m/s within 1.9\,ms by detuning the molasses laser, before the atoms are cooled to $1.0\,\mu$K within 3\,ms. After 44\,ms the atoms reach the rectangular TE$_{401}$ state selection cavity, 18.5\,cm above the cooling region, and the cylindrical TE$_{011}$ Ramsey cavity \cite{Schroeder2002} after 179\,ms, 63.5\,cm above the cooling region. The maximum height of the ballistic trajectory is 1.00\,m and the Ramsey interrogation time is $T_\mathrm{R}=545$\,ms, yielding an 0.92\,Hz FWHM central Ramsey fringe. The detection zone is 30\,cm above the cooling zone. 

Unlike CSF1, the state selection cavity is outside of the magnetically shielded interrogation region, between the cooling and detection zones. This allows pulsing of sufficiently strong magnetic fields to state select using rapid adiabatic passage \cite{PereiraDosSantos2002, Kazda2013} (see subsection~\ref{subsec:Coll}).

The CSF2 laser system is described in \cite{Gerginov2010, Dobrev2016}. Briefly, master laser light from a commercial ECDL injection locks two slave laser diodes that provide the molasses and detection laser light and another slave laser for the LVIS-MOT. Two DBR laser diodes provide repumping light for the cooling and detection zones, and the LVIS-MOT. As in CSF1, the laser light is delivered to the fountain by polarization-maintaining optical fibres. 

The CSF2 repumping laser propagates vertically downward on the fountain axis to the molasses region, entering through a window at the top of the fountain, instead of being superimposed with molasses beams.  This loads atoms from the cold LVIS atom beam closer to the vertical fountain symmetry axis \cite{Dobrev2016}, increasing the number of detected atoms and reducing the distributed cavity phase shift (see subsection~\ref{subsec:DCP}). The combination of the slow-atom beam loading and a vertical repumping increases the number of detected atoms by a factor of more than 40, as compared to molasses loading from a caesium vapor cell \cite{Dobrev2016}. We stabilize the molasses laser intensity with a digital control loop, using liquid crystal variable retarders and polarizers, to reduce drifts of the mean position, shape, and temperature of the atom cloud, and thereby drifts of the distributed cavity phase shift (see subsection~\ref{subsec:DCP}). As in CSF1, commercial laser shutters block the laser light during the microwave interaction phase (see also subsection~\ref{subsec:ACStark}).

\subsection{Common modifications of CSF1 and CSF2}
\label{subsec:common}

Both fountains use new microwave synthesizers based on a shared 9.6\,GHz optically stabilized microwave oscillator (OSMO). It gets its high short-term frequency stability from a common $1.5\,\mu$m cavity-stabilized fibre laser, via a commercial femtosecond comb \cite{Lipphardt2017}. The interrogation frequency is synthesized from the 9.6\,GHz with a divider chain and a Direct Digital Synthesizer (DDS) \cite{Gupta2000,Gupta2007,Kazda2018}. Formerly, we used a series of frequency multiplications and a commercial synthesizer \cite{Schroeder1991}, and both fountains operated autonomously by steering a quartz oscillator to the atomic resonance frequency. The quartz frequencies were then compared to a hydrogen maser using a commercial phase comparator \cite{Weyers2001, Gerginov2010}. The new system locks the OSMO to a hydrogen maser in the long-term. In this way, CSF1 and CSF2 now operate in a non-autonomous mode where each fountain digitally steers the DDS, at $\approx 7.3$\,MHz, and we calculate the frequency difference between the fountains and the hydrogen maser after each fountain cycle. The phase noise of the OSMO contributes little, via the Dick effect, to the best achieved CSF1 and CSF2 instabilities \cite{Santarelli1998,Lipphardt2017}.

Both fountain syntheses chains have electronic switches, which can alternatively select a 9.6\,GHz microwave oscillator stabilized to a low-noise quartz oscillator, which is locked to the hydrogen maser \cite{Gupta2007}. These switches are automatically activated when the fountain control software detects an anomaly in the measured atom number, transition probability or frequency deviations, due to a potential fault in the optically stabilized microwaves. When this occurs, the fountain stabilities are degraded due to the Dick effect.

\subsection{Fountain operation modes}
\label{subsec:opmode}

Both fountains run nearly every day in either the PFS or UTC mode:

\begin{description}

\item[(a)] PFS mode \newline
The Primary Frequency Standard (PFS) mode provides the highest level of accuracy and is used for calibrations of the scale unit of TAI, optical frequency measurements, and other internal and external frequency comparisons. This mode aims for continuous operation for a scheduled measurement period, typically as long as four weeks. To reduce the systematic uncertainty, a number of adjustments and system checks are performed before starting the measurement. These include laser beam adjustments, optimizing injection locks and fibre couplings, and inspections, such as checking that the central Ramsey fringe is used to determine the quadratic Zeeman shift (subsection \ref{subsec:Zeeman}), the microwave powers in the cavities are optimal, and  microwave leakage fields are negligible (subsection \ref{subsec:Leakage}).

When using this mode for measurements, several frequency shifts are evaluated at the end of the measurement period, so that their evaluation can be based on data that is collected in parallel with the actual frequency measurement. The PFS mode encompasses the UTC mode, described next.

\item[(b)] UTC mode \newline
The default is for the fountains to run in the UTC mode to steer UTC(PTB) \cite{Bauch2012}. The steering is sufficiently good when a single fountain contributes at least 6 hours of data each day, although much more data is almost always available. Thus, the fountains can be off line for maintenance or other experimental work, for example, during normal working hours. 

In the UTC mode, the quadratic Zeeman shift (subsection \ref{subsec:Zeeman}), blackbody radiation shift (subsection \ref{subsec:BBR}) and collisional shift (subsection \ref{subsec:Coll}) are evaluated in advance, so that the corresponding corrections are continuously applied, along with the constant relativistic corrections (subsection \ref{subsec:RRS}). Even a relatively large fountain offset of $1\times 10^{-15}$, e.g. due to an errant collisional shift prediction, produces a time scale deviation of only 2.6\,ns after 30 days. 

\end{description}

We note that, in both operation modes, we periodically perform magnetic field measurements (subsection \ref{subsec:Zeeman}) and regularly switch between low and high density operation to measure the collisional shift coefficient (subsection \ref{subsec:Coll}). As a result, CSF1 and CSF2 normally operate with dead times approaching $1-1.5$\%.

\section{Evaluation of systematic effects and uncertainty contributions}
\label{sec:Systematics}

References \cite{Weyers2001, Gerginov2010} reported the first systematic uncertainty evaluations of CSF1 and CSF2, and \cite{Weyers2001b, Weyers2012} reported subsequent improved evaluations. Here we describe the most recent evaluations and the current status of the individual uncertainties.

\subsection{Quadratic Zeeman shift}
\label{subsec:Zeeman}

The nonzero vertical magnetic field $B$ along the atom trajectories above the Ramsey cavity produces a quadratic Zeeman frequency shift of the clock transition frequency. With $B\approx 0.15\,\mu$T in CSF1 and CSF2, this is by far the largest systematic frequency shift, $\approx 1\times 10^{-13}$. Nonetheless, its evaluation is rather straightforward and yields an insignificant uncertainty contribution, well below $10^{-16}$. 

In normal operation for frequency measurements with both fountains, the value of the quadratic Zeeman shift is determined by automated periodic measurements of the frequency detuning $f_{(1,1)}$ of the $\ket{F=4, m_F=1}$ to $\ket{F=3, m_F=1}$ transition from the clock transition frequency $\nu_0$, typically every hour or two, for about 0.5\% of the measurement time. For these measurements, the fountains switch to a mode in which the state selection of $\ket{F=3, m_F=0}$ atoms (see section \ref{sec:description}) is deactivated. Since the magnetic field drifts slowly, this procedure is sufficient to track the magnetically-sensitive transition frequency within the $\approx0.9$\,Hz FWHM of the Ramsey fringes.  The fractional quadratic Zeeman shift of the clock transition frequency is: 

\begin{equation}
\label{eq:Zeeman}
	\frac{\delta \nu_{\mathrm{z}}}{\nu_0} = 8 \left(\frac{f_{\left(1,1\right)}}{\nu_0}\right)^2.
\end{equation}

To ensure that we always measure the central Ramsey fringe of the $\ket{F=4, m_F=1}$ to $\ket{F=3, m_F=1}$ transition, every few months we map the magnetic field along the atomic trajectories in and above the Ramsey cavity, as outlined in \cite{Weyers2000}. To date, we have not observed an incorrect assignment of the central Ramsey fringe in either fountains. 

In the PFS mode, the magnetic field during the measurement campaign is averaged and the resulting correction for the quadratic Zeeman shift is subsequently applied. The $<0.05$\,Hz statistical measurement uncertainty of the $\ket{F=4, m_F=1}$ to $\ket{F=3, m_F=1}$ transition frequency yields an uncertainty for the quadratic Zeeman frequency shift of the clock transition frequency of less than $1.0\times 10^{-17}$. Since the standard deviation of the magnetic field along the atomic trajectories is clearly below 1\,nT, the inhomogeneity of the magnetic field \cite{Weyers2000} contributes well less than $5\times 10^{-18}$ to the uncertainty. This gives an overall uncertainty of the quadratic Zeeman correction in the PFS mode of $<1.0\times 10^{-17}$.

In the UTC mode the correction is updated every few weeks, resulting in frequency errors less than $1\times 10^{-16}$.

\subsection{Blackbody radiation shift}
\label{subsec:BBR}

The second largest frequency shift of CSF1 and CSF2 is caused by the electric field of the ambient blackbody radiation. To evaluate this so-called blackbody radiation shift $\delta \nu_{\mathrm{BBR}}$, we use:

\begin{equation}
\frac{\delta \nu_{\mathrm{BBR}}}{\nu_0} = \frac{k_0 E_{300}^2}{\nu_0} \left(\frac{T}{300\mathrm{K}}\right)^4 \left(1+\epsilon \left(\frac{T}{300\mathrm{K}}\right)^2\right)
\label{eq:BBR}
\end{equation}

\noindent with the ambient temperature $T$ in Kelvin, the RMS electric field of the blackbody radiation at 300 K, $E_{300} = 831.9$\,V/m, and coefficients $k_0 = -2.282(4) \times 10^{-10}$\, Hz/(V/m)$^2$ \cite{Rosenbusch2007} and $\epsilon = 0.013$ \cite{Angstmann2006}. For the latter, an uncertainty of 10\% is assumed \cite{Rosenbusch2007}. 

The ambient temperature in the atomic interaction region is given by three(four) Pt100 resistors along the vacuum tube surrounding the Ramsey cavity of CSF1(CSF2) \cite{Weyers2001,Gerginov2010}. The uncertainty from the Pt100 resistors is $0.11$\,K, and all temperatures are monitored continuously. The temperature stability of the air conditioning of the room housing the fountains yields individual temperature measurements that typically stay within an interval of $\pm 0.2$\,K throughout typical measurement periods lasting several weeks. In the PFS mode, the temperature data for the individual Pt100 resistances during a measurement campaign is first temporally averaged. In a second step, these temperature values are weighted with the dwell time of the atoms in the different regions of the vacuum tube and then averaged. The correction for the blackbody radiation shift is then calculated from (\ref{eq:BBR}). 

In CSF1 the measured temperature gradients along the vacuum tube and the measured maximum temperature difference between the three Pt100 resistors are clearly below $0.3$\,K. The observed gradients are mostly from the MOT-coils heating the lower part of the atomic interaction region. To bound the uncertainty of the measured temperature, we quadratically add the latter value and the Pt100 uncertainty, giving an overall uncertainty of $0.32$\,K. In CSF2, which has no MOT-coils, the measured temperature gradients along the vacuum tube and the measured maximum temperature difference between the four Pt100 resistances are well below $0.2$\,K. After including the Pt100 resistor uncertainty, we obtain a temperature uncertainty of $0.23$\,K.

These temperature measurements lead to frequency corrections of $165.66(80)\times 10^{-16}$ for CSF1 and $165.21(63)\times 10^{-16}$ for CSF2 in the PFS mode, for typical average ambient temperatures of $23.2^\circ$\,C ($23.0^\circ$\,C) for CSF1 (CSF2). In the UTC mode the correction is updated every few weeks, resulting in frequency errors usually less than $1\times 10^{-16}$ for both fountains.

\subsection{Relativistic redshift and relativistic Doppler effect}
\label{subsec:RRS}

For clock comparisons and scale unit measurements of TAI and UTC(PTB), the output frequencies of CSF1 and CSF2 are corrected for the relativistic redshift. The redshift is $(W_0 - W)/c^2$, where $W_0$ is the reference zero gravity potential, and the clock's gravity potential $W$ has to be precisely determined. The relevant height of a fountain clock is the time-averaged height of the atoms between the two Ramsey interactions. 

Under the European EMRP project ``International Timescales with Optical Clocks'' (SIB55 ITOC), the gravity potential was newly determined with respect to the conventional zero potential, $W_0(\mathrm{IERS2010}) = 62\,636\,856.0\,\mathrm{m}^2 \mathrm{s}^{-2}$, at the sites of the European metrology institutes INRIM(Italy), NPL (UK), LNE-SYRTE (France) and PTB (Germany) \cite{Denker2018}. The project used a combination of GPS based height measurements, geometric levelling and a geoid model, refined by local gravity measurements.

With the resulting gravity potentials at local reference markers at PTB, geometric levelling, and accounting for the respective fountain geometries and launch velocities, we determine the frequency corrections for the combined relativistic redshift and the relativistic time-dilation, which depends on the changing atomic velocity during the ballistic flight above the Ramsey cavity. We obtain relativistic frequency corrections of $-85.56(2) \times 10^{-16}$ for CSF1 and $-85.45(2) \times 10^{-16}$ for CSF2, which are applied in both the PFS and the UTC modes. The specified uncertainty is dominated by the uncertainty of the gravity potential at the local reference markers \cite{Denker2018} and is only applicable if all clocks being compared refer to $W_0(\mathrm{IERS2010})$ (see \cite{Guena2017}, e.g.). Because there is presently no exact internationally accepted geoid definition, i.e. an agreed upon zero potential value, we take into account an uncertainty of $3\times 10^{-17}$ (reflecting a height uncertainty of $\approx 0.3$\,m) when CSF1 and CSF2 contribute to TAI. Since both fountains are collocated, the uncertainty of the difference of their relativistic corrections is safely below $1\times 10^{-18}$ for frequency comparisons of CSF1 and CSF2.

\subsection{Collisional shift}
\label{subsec:Coll}

The frequency shift resulting from the cold collisions of the caesium atoms \cite{Tiesinga1992,Gibble1993} is linearly proportional to the cold atom density. The shift is measured by operating the fountain at high and low atom cloud density \cite{Gibble1993,Wynands2005}. The density of the atoms is periodically varied using the state selection cavity, by changing the microwave amplitude in CSF1 \cite{Gibble1993,Wynands2005} and using Rapid Adiabatic Passage (RAP) in CSF2 \cite{ PereiraDosSantos2002,Kazda2013}. For each frequency measurement, the fountains alternately operate at high and low density every few hundred fountain cycles. This yields a differential measurement of the collisional shift, removing the frequency drift of the hydrogen maser reference.

In CSF1 the density, and thereby the number of atoms contributing to the signal, is changed by switching the microwave amplitude in the state selection cavity between a $\pi$-pulse and a $\pi/2$-pulse. Because we use a MOT, the initial atom cloud size is small, $\sigma=0.29$\,mm, and horizontally well-centered on the fountain axis. As a result, the cloud temperature, $T=1.8\,\mu$K, leads to a cloud size in the state selection cavity on the order of $\sigma = 1$\,mm, and the microwave amplitude variation across the cloud is less than 1\% in the rectangular cavity. The small microwave amplitude variation therefore changes cloud size very little between high and low density and yields an accurate density extrapolation. 

\begin{figure}[t]
  \includegraphics[width=\columnwidth]{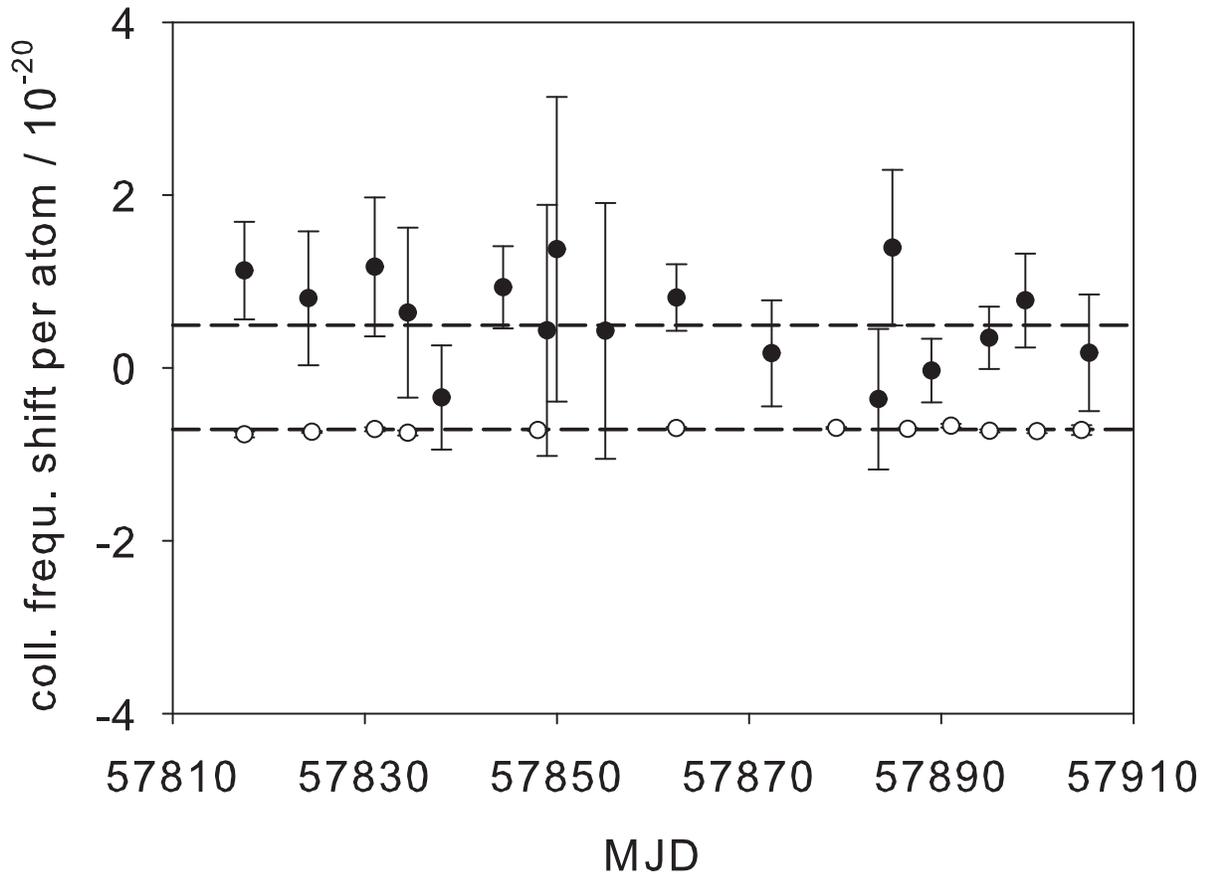}
	\caption{Collisional frequency shift per detected atom for CSF1 (full symbols) and CSF2 (open symbols) over a three months period (MJD: Modified Julian Date).  Dashed lines represent the weighted averages. The positive CSF1 collisional frequency shifts result from lower collision energies due to launching atoms from a MOT \cite{Szymaniec2007a}. The variation of the uncertainties of the individual CSF1 measurements is mostly because the measurement durations ranged from 0.7 to 14 days. The higher frequency stability of CSF2 yields smaller uncertainties of the CSF2 collisional shift coefficients (see section~\ref{sec:Instabil}). Atom numbers were calibrated by signal-to-noise measurements in the quantum projection noise limited regime \cite{Weyers2000}.}
	\label{fig:CollShift} 
\end{figure}

The results of the density extrapolations are collisional shift coefficients (figure~\ref{fig:CollShift}), the collisional frequency shift divided by the detected atom number \cite{Weyers2001,Weyers2001b}. To reduce the statistical uncertainty of the collisional shift coefficients utilized for frequency evaluations, we typically take the average of shift coefficient measurements over three months. For the UTC mode (see subsection~\ref{subsec:opmode}), the applied collisional shift coefficient is regularly updated so that it corresponds to the upcoming measurements. For the PFS mode, the final collisional shift coefficient is calculated after the evaluation interval and then the collisional shift correction is applied.

We calculate the uncertainty of the measured collisional frequency shift from the dominating statistical uncertainty of the collisional shift coefficient and a conservative 10\% systematic collisional shift uncertainty. The latter takes into account long term drifts of the shift coefficients and potential deviations between the actual and the measured density variation, caused by a potentially imperfect proportionality between the actual effective density and the measured number of atoms. Because the processing of the collisional shift coefficients in CSF1 entangles statistical and systematic uncertainties, we attribute an overall collisional shift uncertainty to the systematic uncertainty budget. Since the CSF1 operation parameters are close to the parameters that cancel the collisional shift \cite{Szymaniec2007a}, the measured relative collisional frequency shift is normally less than $10^{-15}$ and its uncertainty is a few parts in $10^{16}$. 

In CSF2 the density is varied using Rapid Adiabatic Passage (RAP) as the atoms traverse the state selection cavity \cite{PereiraDosSantos2002, Kazda2013}. To select all the atoms (full RAP pulse), we apply a 4\,ms pulse with a Blackman amplitude and a 5\,kHz frequency chirp and, for the half RAP pulse, we stop the chirp on resonance, after 2\,ms. These pulses change the local cloud density uniformly by nearly exactly a factor of two, when a full or a half RAP microwave pulse is applied. Thus, the frequency difference between high and low density $f_H-f_L$ gives the collisional shift for low density, and twice the frequency difference is the collisional shift for high density. 

Simulations show that the RAP used in CSF2 effectively reduces the inhomogeneity due to changing the density. The remaining small changes of the cloud size for high and low density lead to a small 0.3\% uncertainty of the collisional shift correction. However, to assess the accuracy of the collisional shift evaluation, we need to account for other effects. First, the broad Fourier spectrum of the half RAP pulse transfers atoms with a small probability from the $\left|4,m_F= 0,\pm1 \right\rangle$ states to the $\left|3,m_F= \pm1\right\rangle$ states \cite{Kazda2013}. The probability depends on the static magnetic field strength and its direction in the state selection cavity, as well as the amplitude of the RAP microwave pulse. The transferred atoms in the $\left|3,m_F= \pm1 \right\rangle$ state are largely unaffected by the subsequent pushing laser pulse and the two Ramsey interactions, but they contribute to the collisional shift and the total detected atom number $N_{tot}$ at low density (half RAP pulse). As a result, the collisional shift evaluation can be distorted and the ratio of the high and low densities may deviate from two. 

Using a simulation of the RAP that includes the $\left|F,m_F=\pm1 \right\rangle$ as well as the $\left|F,m_F=0 \right\rangle$ states, we reduced the number of atoms transferred to the $\left|3,m_F=\pm1\right\rangle$ states. To separate the individual $\left|4,m_F\right\rangle \rightarrow \left|3,m_F\right\rangle$ transitions, the magnetic field in the CSF2 state selection cavity is pulsed to $B^S_z=52.6$\,$\mu$T during the RAP pulses. We use a RAP pulse amplitude of $100\times\Omega_R$, the Rabi frequency for an on-resonance $\pi$ pulse. For these parameters, the simulation shows that the fraction of atoms unintentionally transferred to the $\left|3,m_F=\pm1\right\rangle$ states by the half RAP pulse is $0.1$\% of the state-selected $\left|F,m_F=0\right\rangle$ atoms at high density.
 
Further, the pushing laser light after the RAP pulse non-resonantly optically pumps a small fraction of atoms from the $\left|F=4\right\rangle$ to the $\left|F=3\right\rangle$ ground state \cite{Marion2004,Kazda2013}. The contribution to the collisional shift due to these atoms is almost the same at high and low density and cannot be determined from the frequency difference used to calculate the collisional shift. For half RAP pulses, the off-resonance optical pumping will additionally project part of the $\left|F,m_F=0\right\rangle$ superposition created by the RAP pulse onto the $\left|F=3\right\rangle$ state. The optically pumped population of $\left|3,m_F\neq0\right\rangle$ again distorts the collisional shift evaluation and the high and low density ratio deviates from two.

To ascertain the optically pumped population in the $\left|3,m_F\right \rangle$ states, we periodically turn off the RAP pulse and determine the number of optically pumped atoms $N_{0}$ from the number of atoms in the $\left|F=4\right\rangle$ and $\left|F=3 \right\rangle$ states. (During this measurement, some of the optically pumped $\left|3,m_F=0\right\rangle$ atoms are transferred to the $\left| 4,m_F=0\right\rangle$ state by the Ramsey pulses.) The fraction of optically pumped atoms is typically $0.6$\% of the state-selected $\left| F,m_F=0\right\rangle$ atoms at high density. Independent measurements show these atoms are nearly uniformly distributed over the $\left|3,m_F\right\rangle$ states. 

To determine the systematic error of the collisional shift due to atoms transferred to the $\left|3,m_F\neq0\right\rangle$ states, either by the RAP pulse or the pushing laser, we use the measured collisional shift ratios for individually populated $\left|F=3,m_F\neq0\right\rangle$ states \cite{Marion2004,Papoular2012}. From these ratios we estimate that our collisional shift evaluation results are distorted by less than 0.4\%, where the dominant systematic is due to the optical pumping. This uncertainty includes an estimated uncertainty for the collisional shift ratios because other measurements \cite{Bennett2017} suggest that they may have a strong dependence on the collision energy. With the 0.3\% distortion from residual local variations of the density ratio, we arrive at a total systematic uncertainty of 0.5\% for the collisional shift evaluations of CSF2. 

We regularly check the RAP transfer efficiency by measuring the total number of atoms, $N_{tot}$, and the number of atoms from the off-resonance optical pumping, $N_{0}$. Here, this background is removed, $N'^{\mathrm{HiD}}_{tot}=N^{\mathrm{HiD}}_{tot} - 8/9\times N_{0}$ and $N'^{\mathrm{LoD}}_{tot} = N^{\mathrm{LoD}}_{tot} - 8.5/9\times N_{0}$, for high and low density, assuming $1/9$ of the atoms being in $\left|4,0 \right\rangle$. Deviations of $N'^{\mathrm{HiD} }_{tot}/ N'^{\mathrm{LoD}}_{tot}$ from 2 are below 0.2\%. The corrected atom numbers $N'^{\mathrm{HiD}}_{tot}$ and $N'^{\mathrm{LoD}}_{tot}$ and the frequency difference for high and low density give the collisional shift coefficients, as in Fig.~\ref{fig:CollShift}. Multiplying the previously measured collisional shift coefficients by the corrected total atom number gives the collisional shift in the UTC mode, as in CSF1.

To correct the collisional frequency shift of CSF2 in the PFS mode, we use only the high and low density frequency measurements from the same evaluation period to obtain the corrected frequency $2 f_L-f_H$. The measured fractional collisional frequency shift is typically a few parts in $10^{15}$, with a systematic uncertainty of a few parts in $10^{17}$ (0.5\%). Unlike in CSF1, the statistical uncertainty of the collisional shift correction is included in the overall statistical uncertainty, and it dominates (see section~\ref{sec:Instabil}).

\subsection{Distributed cavity phase shift}
\label{subsec:DCP}

Phase gradients in the cylindrical Ramsey cavity produce frequency shifts because the atoms traverse the cavity at different transverse positions on their upward and downward cavity passages. The evaluation of the distributed cavity phase (DCP) shifts is based on the theory developed in \cite{Li2004,Li2010}, which was experimentally verified in \cite{Guena2011}, with further corroboration in \cite{Weyers2012,Li2011}. The last DCP evaluation of CSF2 was reperformed after slow atom beam loading of the optical molasses was implemented \cite{Dobrev2016}.


The theory decomposes the effective transverse phase variations into a Fourier series $cos(m\phi)$, with cylindrical coordinates ($\rho,\phi,z$), and only the $m\le 2$ terms are significant. The $m=0$ phase variations are caused by power flow from the two cavity feeds in CSF1 and CSF2 at the cavity midplanes towards the endcaps and the $m=1$ and $2$ phase variations are caused by transverse power flow from the feeds to the cavity walls. Typically the $m=0$ and $2$ phase variations are sufficiently small that they can be accurately calculated by taking into account the cavity geometry and the sizes and positions of the atom cloud during the cavity passages. In contrast, the $m=1$ transverse phase variations can be large as they depend on precise balancing of the amplitudes of opposing cavity feeds. Unknown resistance inhomogeneities of the copper cavity walls also produce similar power flow and transverse phase gradients. The $m=1$ phase variations are linear gradients near the cavity axis, which can be experimentally evaluated by intentionally increasing the mean transverse displacement of the atom cloud on the two cavity passages \cite{Guena2011,Li2011,Weyers2012}, by tilting the entire fountain or by varying the atom launch-direction. 

The same cavity design is employed in CSF1 and CSF2. Power is supplied via a single cable to a curved waveguide, which in turn supplies power to the inner cylindrical cavity via two opposing slits \cite{Schroeder2002}. In contrast to other cavity designs with independent cavity feeds, this cavity design precludes experimentally increasing the $m=1$ DCP shift by alternately supplying one feed or the other as the fountains are tilted \cite{Guena2011,Li2011}. As a consequence, in CSF1 and CSF2 this technique cannot be utilized to adjust the feed balance or to align the fountain to be vertical, and therefore eliminate $m=1$ DCP shifts. Instead, we developed experimental methods to measure the mean transverse atom cloud position in the Ramsey cavity. We use the microwave $\ket{F=3, m_F=0}$ to $\ket{F=4, m_F=1}$ transition probabilities with frequency detunings and varying fountain tilts or atom launch directions \cite{Nemitz2012,Weyers2012}. The resulting atom cloud positions and their uncertainties are also used to calculate the $m=0,2$ DCP shifts and uncertainties, for which we use finite element calculations of the cavity fields \cite{Li2004,Li2010,Weyers2012}. 

For CSF1 we tilt the entire fountain to vary the horizontal cloud position in the cavity, whereas in CSF2, the cooling zone is swivel-mounted \cite{Gerginov2010} so that the launch direction can be varied without moving the rest of the fountain. Comparing these two techniques, the first case requires larger tilt angles to achieve the same cloud displacement. For CSF1, we get a negligible initial cloud offset with an uncertainty of $0.5$\,mm. However, using optical molasses loaded from the LVIS in CSF2, instead of a MOT, not only produces a larger initial cloud of $\sigma = 2.5$\,mm, but also an initial cloud offset of $2.5(0.7)$\,mm in the direction of the LVIS setup. This significant offset is due to the asymmetric loading from the LVIS: the slow atoms are quickly decelerated by the molasses beams before they reach the center of the molasses zone. 

For normal operation, both fountains are vertically aligned to maximize the detected atom number, i.e. to horizontally center the falling atom cloud on the Ramsey cavity apertures. We next describe the determination of the individual DCP frequency shift contributions and their uncertainties for CSF1 and CSF2.

\subsubsection{$m=0$ DCP frequency shifts}
\label{subsubsec:DCPm=0}
The power flow from the cavity feeds at the cavity midplane to the endcaps produces large longitudinal phase variations. These yield small frequency shifts at optimum microwave amplitude, two Ramsey $\pi/2$ pulses, and large frequency shifts at elevated microwave amplitudes, e.g. $4.25\,\pi/2$ and $8.25\,\pi/2$ pulses \cite{Li2005,Li2010}. These results explain our previously observed frequency shifts for CSF1 and CSF2 at elevated microwave amplitudes \cite{Weyers2007,Gerginov2010,Gerginov2010a}. We therefore no longer include  uncertainties related to microwave power dependence in our error budgets. 

The most significant $m=0$ DCP uncertainty arises from potentially different electrical conductivities of the top and bottom cavity endcaps. With an upper limit of 20\% \cite{Weyers2012}, the $m=0$ DCP frequency shift is $+0.09(17)\times 10^{-16}$ for CSF1. For CSF2, the larger cloud size gives a better cancellation of the longitudinal phase variations during the fountain ascent and descent of the atoms through the Ramsey cavity, leading to a smaller shift of $+0.035(0.020)\times 10^{-16}$.

\subsubsection{$m=1$ DCP frequency shifts}
\label{subsubsec:DCPm=1}
The dominant DCP uncertainty in both CSF1 and CSF2 is from the $m=1$ DCP frequency shifts due to transverse phase gradients. To experimentally evaluate this shift, we first measured the fountain frequencies for two opposing fountain tilts (CSF1) or atom launch directions (CSF2) in two orthogonal directions. For CSF1 the \textit{x } tilt axis is at an angle of $22^\circ$  to the cavity feed axis, whereas, for CSF2, the \textit{x } tilt direction coincides with the cavity feed axis. The observed cloud offset in CSF2 is $(x,y)=(1.8(0.7)\,\mathrm{mm},-1.8(0.7)\,\mathrm{mm})$, where the origin is the longitudinal fountain symmetry-axis. Along each axis, a number of $\sim24$-hour measurements of the fountain frequency were performed, alternating between positive and negative tilts. The other fountain, or the electric quadrupole transition frequency of a single trapped $^{171}$Yb$^+$ ion \cite{Tamm2014}, was used as a frequency reference.

\begin{figure}[t]
  \includegraphics[width=\columnwidth]{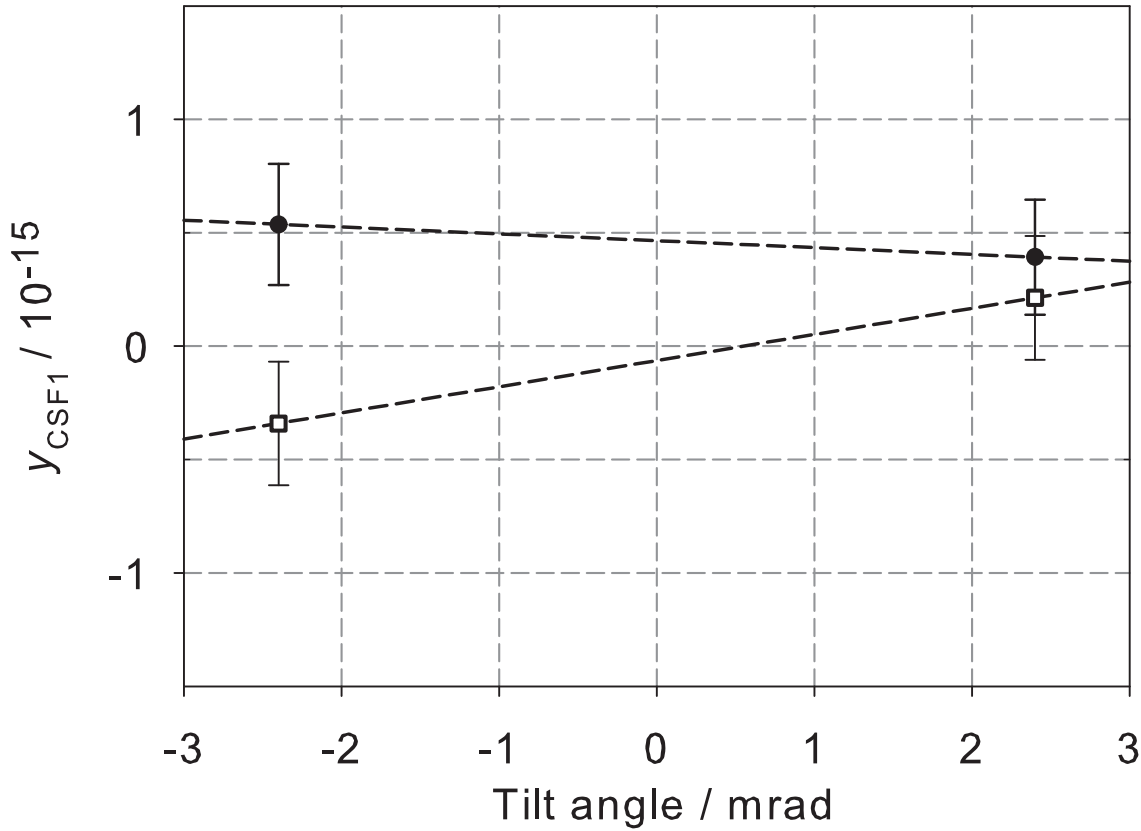}
	\caption{Frequency of PTB-CSF1 versus fountain tilt, $\pm2.4$\,mrad at $22^\circ$ (full circles) and $112^\circ$ (open squares) from the feed direction. The fits (dashed lines) are $-0.30(76)\times 10^{-16}\,\mathrm{mrad}^{-1}$ and $1.15(1.05)\times 10^{-16}\,\mathrm{mrad}^{-1}$. CSF1 normally operates at zero tilt angle, which simultaneously maximizes the number of detected atoms and minimizes $m=1$ DCP frequency shifts.}
	\label{fig:CSF1Tilt} 
\end{figure}

\begin{figure}[t]
  \includegraphics[width=\columnwidth]{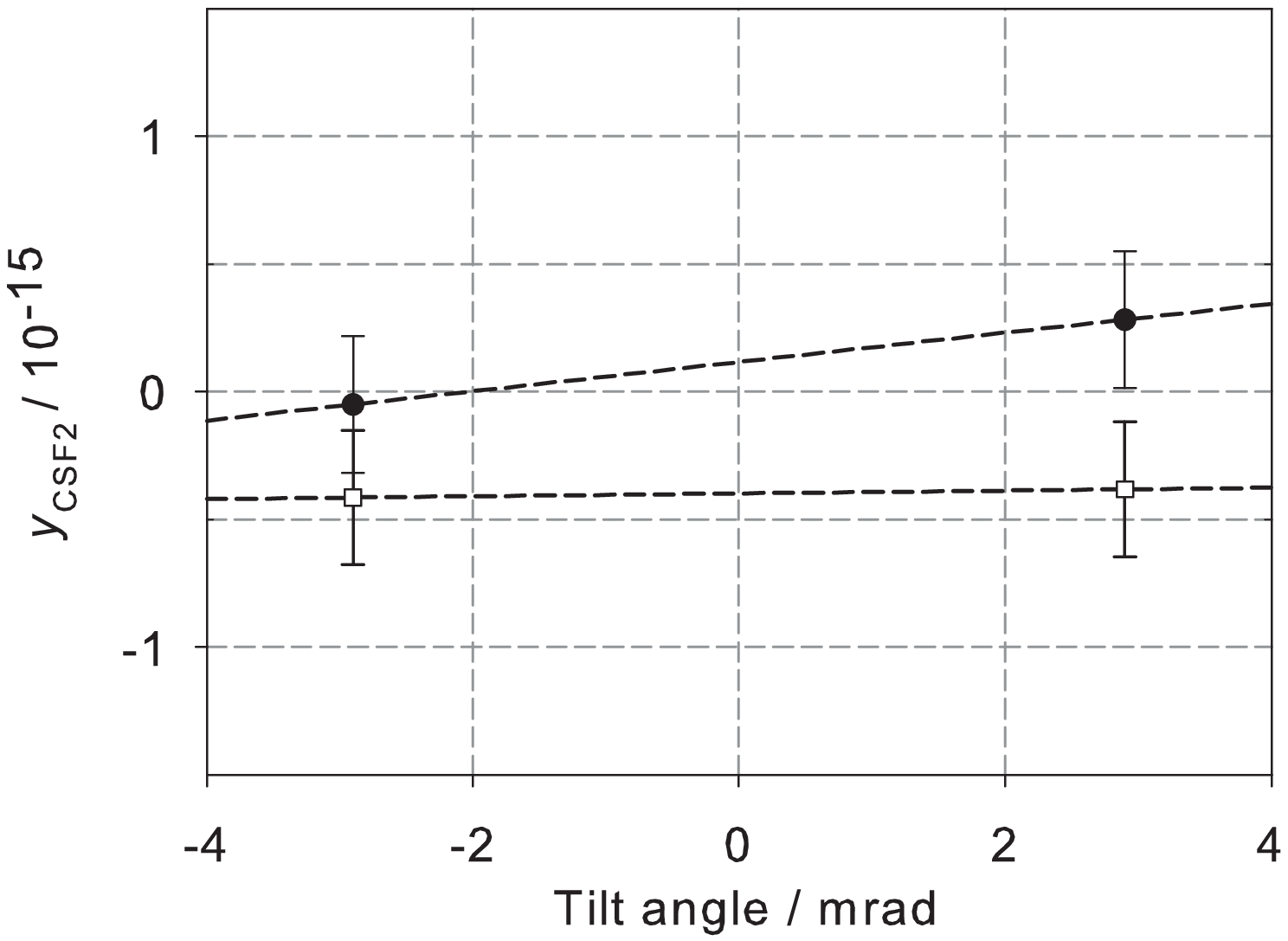}
	\caption{Frequency of PTB-CSF2 versus launch angle, $\pm2.9$\,mrad parallel (full circles) and perpendicular (open squares) to the feeds. The fits (dashed lines) are $0.57(66)\times 10^{-16}\,\mathrm{mrad}^{-1}$ and $0.06(64)\times 10^{-16}\,\mathrm{mrad}^{-1}$. CSF2 normally operates at zero tilt to maximize the number of detected atoms, which gives a non-zero $m=1$ DCP frequency shift due to the initial offset of the atom cloud from the fountain axis.}
	\label{fig:CSF2Tilt} 
\end{figure}

Figure~\ref{fig:CSF1Tilt} shows the shifts of CSF1 for fountain tilts of $\pm 2.4$\,mrad along both axes. The tilt sensitivities are $-0.30(76)\times 10^{-16}\,  \mathrm{mrad}^{-1}$ along the x-axis and $1.15(1.05)\times 10^{-16}\,\mathrm{mrad}^{-1}$ along the y-axis. Figure~\ref{fig:CSF2Tilt} shows the similar shifts of CSF2 for atom launch directions changed by $\pm 2.9$\,mrad, which yields tilt sensitivities of $0.57(66)\times 10^{-16}\,\mathrm{mrad}^{-1}$ along x and $0.06(64)\times 10^{-16}\,\mathrm{mrad}^{-1}$ along y. All of the measured phase gradients are consistent with zero. 

From these tilt measurements, the $m=1$ DCP frequency shift of CSF1 is zero with uncertainties of $0.41\times 10^{-16}$ (x-direction) and $0.78\times 10^{-16}$ (y-direction). For CSF2, to account for the initial cloud offset, we simulated the ballistic flight of the atom cloud to determine the launch angle that maximizes the number of detected atoms from true vertical, for which the atom cloud has the same mean transverse position on the two cavity passages and therefore gives no $m=1$ DCP shift \cite{Weyers2012}. The launch angles are $+1.4(5)$\,mrad for the x-axis and $-1.4(5)$\,mrad for y. Combining these with the measured tilt sensitivities, the $m=1$ DCP frequency shifts are $+0.80(1.02)\times 10^{-16}$ for x tilts and $-0.08(95)\times 10^{-16}$ for y. The total $m=1$ DCP frequency shift is then $+0.72(1.39)\times 10^{-16}$, which is larger than for CSF1 due to the relatively large initial cloud offset and the uncertainties of the tilt sensitivities.

\subsubsection{$m=2$ DCP frequency shifts}
\label{subsubsec:DCPm=2}
The $m=2$ DCP shifts, from quadrupolar phase variations, vanish for a horizontally symmetric cloud that is launched vertically and detected homogeneously \cite{Li2010}. We calculate this shift using the specific fountain parameters and geometry, including the detection zone geometry and its orientation with respect to the cavity feeds. The DCP shift from detection inhomogeneities is suppressed when the feeds are at $45^\circ$, halfway between the perpendicular detection laser propagation axis and the imaging axis \cite{Li2010}. In CSF1 the detection laser direction is $22^\circ$ from the feeds and, in CSF2, $37.5^\circ$. The $m=2$ DCP shifts are therefore suppressed by factors of 1.4 and 3.9. We calculate $m=2$ DCP frequency shifts of $-0.05(24)\times 10^{-16}$ for CSF1 and $-0.48(60)\times 10^{-16}$ for CSF2. The shift is larger for CSF2 due to its initial cloud offset. The uncertainties include the cloud position uncertainties and the modeled detection inhomogeneities.

\subsubsection{DCP frequency shift summary}
\label{subsubsec:sumDCP}
The $m=0,1,2$-DCP shifts and uncertainties are summarized in Table~\ref{tab:DCP}. In comparison to the previous DCP evaluation of CSF2 \cite{Weyers2012}, the largest difference is an increase of the $m=2$ DCP shift and its uncertainty, due to the larger initial offset of the atom cloud loaded from the LVIS \cite{Dobrev2016}. The DCP corrections of Table~\ref{tab:DCP} are applied for both the PFS and the UTC modes.

\begin{table}
\caption{\label{tab:DCP} Individual DCP frequency shifts of PTB-CSF1 and PTB-CSF2 and their uncertainties (parts in $10^{16}$).}
\begin{indented}
\lineup

\item[]\begin{tabular}{@{}lllll}
\br
& \multicolumn{2}{c}{\bfseries CSF1} & \multicolumn{2}{c}{\bfseries CSF2} \\
\ns\ns
&\crule{2}&\crule{2}\\
Effect & Shift & Uncertainty & Shift & Uncertainty \\
\br

DCP $m=0$                  									&	$+0.09$	  &		0.17	& $+0.035$	& 0.020\\
DCP $m=1$	(x-axis)				             			&	$\m0.00$  &		0.41	& $+0.80$		& 1.02\\
DCP $m=1$	(y-axis)													&	$\m0.00$  & 	0.78	&	$-0.08$		& 0.95\\
DCP $m=2$      						              		&	$-0.05$	  &		0.24	& $-0.48$   & 0.60\\
\mr
total                                   		& $+0.04$	  &		0.93	& $+0.28$		& 1.52\\
\br
\end{tabular}
\end{indented}
\end{table}

\subsection{Microwave lensing}
\label{subsec:MWL}

The transverse variation of the amplitude of the microwave field in the Ramsey cavity produces well-known resonant dipole forces \cite{Bloom1962,Sleator1992}. These act as positive and negative lenses on the atomic dressed states, which are subsequently detected either at a positive or negative frequency detuning of the central Ramsey interrogation, thereby yielding a frequency shift \cite{Gibble2006}. Here, in the microwave lensing regime, the atomic wavepackets are restricted by the cavity apertures to be smaller than the microwave wavelength. If the microwave wavelength decreases so that it is much shorter than the size of the atomic wavepackets, the microwave lensing shift then smoothly connects to the photon-recoil frequency shift \cite{Gibble2016}.

To calculate the microwave lensing frequency shifts of CSF1 and CSF2, we use their specific fountain parameters and geometries, as for the DCP calculations. The apertures that clip the atom cloud play a central role in the microwave lensing shift and, for CSF2, two detection apertures contribute in addition to the usual two restrictive apertures in most fountain clocks \cite{Li2011,Gibble2006,Gibble2014,Peterman2016}. 

We derive the microwave lensing frequency shift, including additional detection apertures, in appendix\,A. Near optimal microwave amplitude, $b_1 \approx 1$, we get a very good approximation \cite{Li2011, Gibble2006,Gibble2014,Peterman2016} to the full expression (\ref{eq:Lensing2}) if we neglect the small variation of the state detection in CSF1 and CSF2: 

\begin{eqnarray} 
\label{eq:SimpLens} 
\fl\frac{\delta\nu}{\nu_0}  & \approx &  \frac{\lambda}{{8\left({t_2-t_1}\right)}} \frac{b_1 \eta}{\sin\left( {b_1 \eta \frac{\pi}{2}} \right)} \left[ {a_2\left({t_{2L} -t_1}\right) \int_{r_{1L}<a_1}{\int_0^{2\pi} {\vphantom{\left.  {\frac{\left({t_{1L}} \right)}{\left({t_{2L}} \right)}} \right|_{r_{2L0}}}    J_1\left({k r_1}\right) n_0\left({\vec r_{1L},\vec r_{2L0}} \right)}} } \right. \nonumber\\ 
\fl&& \times\left.  { \Theta \left({a_d-\left|{x_{d0}} \right|} \right) \frac{r_{2L0} \left({t_1-t_{1L}} \right)+r_{1L} \left({t_{2L}-t_1} \right)\cos \left({\phi_{2L0}}\right)} {r_1 \left({t_{2L}-t_{1L}} \right)}} \right|_{r_{2L0}=a_2}\!\!\!\!\mbox{d}\phi_{2L0} \mbox{d}^2 r_{1L} \nonumber\\
\fl&&{}+ \left({t_d-t_1}\right)\frac{t_{2L}-t_{1L}}{t_d-t_{1L}}\sum\limits_\pm {\pm \int_{r_{1L}<a_1}{\int_{-a_2}^{a_2}   {\frac{x_1}{r_1}J_1\left({k r_1}\right)  n_0 \left({\vec r_{1L},\vec r_{2L0}} \right) }}} \\
\fl&& \times \left. \left. \Theta \left({a_2- r_{2L0}} \right) \right|_{x_{2L0} =\frac{\pm a_d \left({t_{2L}-t_{1L}} \right) + r_{1L}\cos\left( {\phi_{1L}} \right)\left({t_d-t_{2L}} \right)}{t_d-t_{1L}}} \mbox{d}y_{2L0} \mbox{d}^2 r_{1L} \vphantom{\int_{r_{1L}}{\int_0^2 {\left.  \frac{\left({t_{1L}} \right)}{\left({t_{2L}} \right)} \right|_{r_{2L0}} }}}\right] \bigg/ \nonumber\\
\fl&&\quad \int_{r_{1L}<a_1}{\int_{r_{2L0}<a_2}{\!\!\!\! n_0 \left({\vec r_{1L},\vec r_{2L0}} \right)\Theta \left({a_d-\left|{x_{d0}} \right|}\right) \mbox{d}^2 r_{2L0}} \mbox{d}^2 r_{1L}}\;\mbox{,} \nonumber 
\end{eqnarray}

\begin{eqnarray*} 
\fl\mbox{where}&\nonumber\\
\fl n_0 \left({\vec r_{1L} ,\vec r_{2L0}} \right) & = n_{00}\: \exp\left[-\frac{{\left| {\vec r_0-\vec r_{00}} \right|^2}}{{w_0^2}}-\frac{{\left|{\vec v-\vec v_0}\right|^2}} {{u^2}}\right]\Theta \left({a_1- r_{1L}} \right)} \qquad{\vec v = \frac{{\vec r_{2L0}-\vec r_{1L}}}{{t_{2L}-t_{1L}}}\\
\fl\hspace{1.5cm}\vec r_\beta & = \frac{{\vec r_{2L0} \left({t_\beta-t_{1L}} \right)+\vec r_{1L} \left({t_{2L}-t_\beta} \right)}}{{t_{2L}-t_{1L}}}}\qquad{\beta\in \left\{{0,1,20,2L0,d0} \right\}\\
\fl\hspace{1.5cm}x_\beta & = \frac{{r_{2L0}\cos\left({\phi_{1L}+\phi_{2L0}}\right)\left({t_\beta-t_{1L}} \right)+ r_{1L} \cos \left({\phi _{1L}} \right)\left({t_{2L}-t_\beta}\right)}}{{t_{2L}-t_{1L}}}} \qquad{\beta  \in \left\{{1,d0} \right\} \mbox{.}  
\end{eqnarray*}

\noindent Here, the integrations are over the transverse positions $\vec r_{1L,2L0}$ at the upward and downward circular apertures \cite{ Gibble2014,Peterman2016}, which have radii $a_{1,2}=5\,\mathrm{mm}$. The atoms pass through these apertures at times $t_{1L,2L}$, are detected at $t _{d0}$, and experience Ramsey pulses at $t_{1,2}$. The $1/e$ velocity halfwidth is $u$, $w_0$ is the initial $1/e$ cloud radius, $n _{00}$ is the peak atomic density, $v_0$ and $r_{00}$ are the initial transverse velocity and position offsets, $\eta =1.120$ for a $5\,\mathrm{mm}$ aperture \cite{Li2010}, $\lambda = h/ mc$, $k=2\pi\nu_0/c$, and the Heaviside functions $\Theta\left({a_{1,2}-r_{\beta}}\right)$ describe the circular cavity apertures at $t_{1L,2L}$ and $\Theta\left({a_d-\left|{x_{d0}} \right|}\right)$, the rectangular detection aperture of halfwidth $a_d=5(7)\,\mathrm{mm}$ for CSF2(CSF1). For CSF1, the second term in $\left\{\ldots  \right\}$ evaluates to zero because its larger detection aperture does not clip the atoms' fluorescence. 

Equation (\ref{eq:SimpLens}) gives microwave lensing frequency shifts of $0.44\times 10^{-16}$ for CSF1 and $0.67\times 10^{-16}$ for CSF2. The full expression (\ref{eq:Lensing2}) yields corrections that are less than $3\times 10^{-19}$ at optimal amplitude for both fountains, with negligible uncertainties from the parameters. Equations (\ref{eq:SimpLens}) and (\ref{eq:Lensing2}) do not include frequency shifts from the small differential phase shifts of the dressed states or from the usually negligible dipole forces during the second Ramsey interaction \cite{Gibble2014,Peterman2016}. These are both normally less than $0.1\times 10^{-16}$ for fountains \cite{Gibble2014,Peterman2016} and we therefore assign an uncertainty of $<0.20\times 10^{-16}$ for the microwave lensing shifts of CSF1 and CSF2. The microwave lensing corrections are applied in both the PFS and UTC modes.

\subsection{AC Stark shifts}
\label{subsec:ACStark}

Nearly resonant laser light during the Ramsey interrogation time produces an AC Stark or light shift of the clock transition \cite{Vanier1989}. We use three techniques to suppress AC Stark shifts during the interrogation time \cite{Weyers2001,Gerginov2010}. First and most important, as mentioned in section~\ref{sec:description}, mechanical shutters block laser light from entering the optical fibres that deliver the laser light to the fountains. Second, RF switches turn off the RF to the acousto-optical modulators (AOMs) that deflect laser light into the fibres. Third, a mechanical shutter blocks the master laser beam that injection-locks the slave laser diodes. The free-running slave lasers have a detuning of $\sim 1$\,nm, strongly reducing any potential AC Stark shifts from the slave lasers. 

We extensively investigated potential AC Stark shifts in both fountain clocks by measuring the fountain frequency while inhibiting individual shutters as in \cite{Weyers2001,Gerginov2010}. The maximum observed shifts are of order $10^{-14}$ and, with shutter extinction ratios greater than $10^{6}$, we conclude that AC Stark shifts are safely below $10^{-18}$, which we take as the uncertainty.

\subsection{Rabi and Ramsey pulling}
\label{subsec:RR}

Rabi and Ramsey pulling, the pulling of the clock frequency by off-resonant excitation of nearby transitions, were recently reevaluated for CSF1 and CSF2 \cite{Gerginov2014}. That work showed that asymmetrically excited Zeeman coherences can potentially enhance the Ramsey pulling. The excitation of the $\ket{F=3, m_F=\pm 1}$ states in CSF1 is kept low and symmetric by ensuring that the microwaves feeding the state selection cavity have no detuning. In CSF2 potential asymmetric coherences from the half RAP pulse (see subsection~\ref{subsec:Coll}) are suppressed by averaging out the phase of the Zeeman coherences of individual atoms. The distribution of atomic transit times and a static magnetic field ($>10\,\mu$T) between the state selection and the Ramsey cavity, which is also inhomogeneous near the magnetic shield apertures (see figure~\ref{fig:CSF1CSF2}), randomizes the phases of the coherences. 

Majorana transitions \cite{Majorana1932,Bauch1993} can excite Zeeman coherences and these can contribute to Ramsey pulling if the coherences are asymmetric or the detection of the $\ket{F, \pm m_F}$ states is inhomogeneous \cite{Gerginov2014,Wynands2007}. Majorana transitions after the state selection and before the Ramsey interrogation, or after the Ramsey interrogation and before the detection can contribute. In a fountain, state selecting atoms in $\ket{F, m_F=0}$ avoids asymmetric coherences as the Majorana transitions can only produce symmetric coherences. In CSF1 and CSF2 we avoid magnetic field reversals and zeroes along the atomic trajectories. Compensation and supplementary coils along the atomic flight path smoothly increase the vertical magnetic field outside of the shielded interrogation region, reducing Majorana transitions to a negligible level \cite{Weyers2001,Gerginov2010,Wynands2007}. 

For both CSF1 and CSF2, the measured asymmetry of the $\left|3,m_F\pm1\right\rangle$ populations is 0.25\% of the $\left|3,0\right\rangle$ population and gives a Rabi and Ramsey pulling uncertainty of $1.3 \times 10^{-18}$ for both fountains \cite{Gerginov2014}.

\subsection{Microwave leakage}
\label{subsec:Leakage}

Unintended microwave fields, beyond the intended Ramsey pulses, after the state selection and before the final state detection can produce frequency shifts~\cite{Boussert1998,Weyers2006,Shirley2006}. Such disturbing fields can arise from microwaves leaking from the microwave synthesizer and entering the flight region through the viewports, or from microwave leakage from the Ramsey or state selection cavities. To reduce these frequency shifts, we implement different methods for our two fountains. 

For CSF1, an interferometric switch ~\cite{Santarelli2009} supplies 407.3\,MHz to generate the 9.2\,GHz microwaves delivered to the Ramsey cavity. The interferometric switch is regularly proven to attenuate 9.2\,GHz by at least $35$\,dB when the atoms are outside the Ramsey cavity, suppressing leakage shifts by at least a factor of $50$ \cite{Weyers2006}. Similarly, when the atoms are outside of the state selection cavity, the amplitude of its 9.2\,GHz microwaves is attenuated by at least $200$\,dB by an electronic attenuator and switch.  

For CSF2 the Ramsey cavity microwaves are instead frequency detuned when the atoms are outside the Ramsey cavity \cite{Lin2009,Kazda2016a}. The DDS in the microwave synthesizer (see subsection~\ref{subsec:common}) phase-continuously steps the frequency by about 390\,kHz and additional phase synchronization preserves the phase of the interrogating microwaves \cite{Kazda2018}. The RAP state-selection microwaves (see subsection~\ref{subsec:Coll}) are detuned by several Megahertz and attenuated by $80$\,dB when the atoms are outside of the state selection cavity. 

With no microwave attenuation, we use a horn antenna as receiver around the fountains and detect no leakage at the sensitivity limit of our spectrum analyzer, $-154$\,dBm in a 1\,Hz bandwidth. From \cite{Weyers2006}, we conclude that the attenuation and detuning reduce the microwave leakage shifts to well below $1 \times 10^{-18}$. 

Potential phase shifts between the two Ramsey interactions from the interferometric switching and frequency detuning are analyzed in the next section, and an associated uncertainty is assigned.

\subsection{Electronics}
\label{subsec:Electr}

Here we discuss effects attributable to the electronic systems that generate and control the microwaves for the fountains. More detail about the microwave synthesis are described in \cite{Gupta2007,Kazda2018,Kazda2018a}.

The phase noise of the Ramsey interrogation signal close to the carrier determines the degradation of the fountain frequency stability due to the Dick effect \cite{Santarelli1998}. Using a heterodyne mixing technique and a cross-spectrum analyzer, we measure the single sideband phase noise of the microwave synthesizer. At 1\,Hz offset frequency, the phase noise relative to the 9.2\,GHz carrier is about $-100$\,dBc/Hz and drops to $-118$\,dBc/Hz at a 10\,Hz offset. This noise is at or below the phase noise of the optically stabilized microwaves and therefore does not limit fountain frequency instabilities above $1 \times 10^{-14} (\tau/1\,\mathrm{s})^{-1/2}$ \cite{Lipphardt2017}.

We use the same phase measurement system to measure the long-term phase stability of the microwave synthesizer. The Allan standard deviation between two synthesizers is $2 \times 10^{-15} (\tau/1\,\mathrm{s})^{-1/2}$. With the measured temperature sensitivity of the synthesizer of less than 1\,ps/K, and the typically slow and periodic environmental temperature variations of 0.2\,K peak-to-peak, which have an $\approx 11000$\,s period, the synthesizers frequency instability is almost one order of magnitude below the lowest CSF2 instability.  

Spectral impurities in the Ramsey interrogation microwaves can produce systematic frequency shifts \cite{Ramsey1955,Shirley1963,Audoin1978}. Such impurities can be introduced by dividers, amplifiers, direct digital synthesis chips in the microwave synthesizer and can be carrier sidebands at multiples of the line frequency. Asymmetric sidebands lead to frequency shifts that depend on the sideband offset from the carrier, the sideband power, and the asymmetry \cite{Audoin1978,Levi2006a}. In our case, all relevant sidebands are within a few kHz of the carrier and are symmetric to within 1\,dB. Phase modulation at the 50\,Hz line frequency produces seemingly symmetric sidebands at -67\,dBc. From the theory in \cite{Audoin1978}, a single 50\,Hz sideband at this level yields a frequency shift of only $4 \times 10^{-18}$. As the contribution from sidebands farther from the carrier frequency is more than an order of magnitude lower, we take this as the maximum uncertainty contribution from all asymmetric sidebands.

Symmetric sidebands can also cause frequency shifts if the sideband frequency is coherent with the fountain cycle ~\cite{Shirley2009,Heavner2006}. Considering the sideband modulation periods of our fountains, we have adjusted the cycle times to average out these frequency shifts after multiples of 200 fountain cycles \cite{Kazda2018,Kazda2018a}. In CSF1 we chose a cycle time of $T_c=1.1155$\,s and for CSF2, $T_c=1.2345$\,s. These provide atom numbers and densities that yield reasonable compromises between the collisional shift uncertainties and the frequency instabilities (see subsections~\ref{subsec:Coll} and \ref{sec:Instabil}).

We also use a phase transient analyzer \cite{Kazda2016} to investigate potential phase perturbations of the Ramsey microwave signal between the two Ramsey interactions of the atoms. These measurements can detect sub-microradian phase drifts or jumps between the two Ramsey interactions, corresponding to frequency shifts at the $10^{-17}$ level. Phase variations may be induced by electronic switching that is synchronous with the fountain cycle, such as by the interferometric switch and the electronics for the phase-preserving frequency detuning (see subsection~\ref{subsec:Leakage}). For both fountains, our measurements bound frequency shifts caused by phase perturbations to $1\times10^{-17}$, which we use as the overall electronics uncertainty.

\subsection{Background gas collisions}
\label{subsec:BG}

The frequency shift due to collisions of cold clock atoms with room temperature background gas atoms can be bounded by measurements of the collision-induced loss of atoms from the clock states \cite{Gibble2013}. From the current reading of the top ion pumps, we obtain for CSF1 a residual gas partial pressure in the low $10^{-7}$\,Pa range and in the mid $10^{-8}$\,Pa range for CSF2. In the cooling zone in the lowest part of the vacuum system of CSF1, caesium atoms are loaded into the MOT from the background caesium vapour (see figure~\ref{fig:CSF1CSF2}). The source of the background caesium is a reservoir, attached to the cooling zone and held at $\approx 285$\,K. The cooling zone and the detection zone above are connected by a tube whose inner diameter is restricted to 1\,cm by a graphite insert to getter ascending background caesium atoms. Thus, a column of room temperature caesium atoms ($\approx 1$\,cm in diameter) remains along the symmetry axis of the fountain, which give the dominant frequency shift due to background gas. In contrast, in CSF2 the atomic beam from the LVIS exits through an 0.5\,mm aperture, which provides differential pumping between the two vacuum zones \cite{Dobrev2016}. Therefore, the background gas frequency shifts of CSF2 arise mainly from collisions of the cold caesium atoms with hydrogen molecules.

From atom loss measurements at different caesium vapour pressures we estimate a $<20$\% atom loss from caesium background gas collisions for the normal operation of CSF1. To estimate the atom loss caused by collisions with hydrogen molecules in CSF2, we turned off and on the top ion-getter pump and recorded the atom loss for different vacuum pressures. When the pump is on, we estimate a 5\% loss for normal conditions. The estimated relative losses $\Delta A$, give a background gas induced fractional frequency shift: 

\begin{equation}
\frac{\delta \nu_\mathrm{{BG}}}{\nu_0} = -\frac{\Delta A}{13.8 \pi \nu_0 T_\mathrm{R}} \; \frac{\Delta C_6}{C_6},
\label{eq:BG}
\end{equation}

\noindent where we assume $\Delta C_6/C_6=1/25000$ for collisions with background caesium atoms and $\Delta C_6/C_6=1/34000$ for collisions with background hydrogen molecules \cite{Gibble2013}. This bounds the frequency shifts due to background gas collisions to $4\times10^{-17}$ for CSF1 and $1\times10^{-17}$ for CSF2. We take these values for the respective systematic uncertainties.

\subsection{Summary of systematic frequency shifts and uncertainties of CSF1 and CSF2}
\label{subsec:Sum}

Table~\ref{tab:uncertainties} summarizes the systematic frequency shifts and uncertainties described in section~\ref{sec:Systematics}. The uncertainty of CSF1 is limited by the statistical uncertainty of the collisional shift evaluation, while the corresponding statistical component of CSF2 is not included in the systematic collisional shift uncertainty but in the overall statistical uncertainty. For CSF2, the systematic uncertainty is dominated by the distributed cavity phase shift, because of the initial cloud offset and the uncertainties of the tilt sensitivities.

\begin{table}
\caption{\label{tab:uncertainties} Uncertainty budgets of CSF1 and CSF2 in the PFS mode: Systematic frequency shift, applied frequency correction, and uncertainty, in parts in $10^{16}$.}
\begin{indented}
\lineup

\item[]\begin{tabular}{@{}lllll}
\br
& \centre{2}{\bfseries CSF1} & \centre{2}{\bfseries CSF2}\\
\ns\ns
&\crule{2}&\crule{2}\\
Systematic frequency shift & Correction & Uncertainty & Correction & Uncertainty\\
\br
Quadratic Zeeman shift                  & $-1079.20^{\rm a}$       & 0.10    & $-998.62^{\rm a}$      & 0.10\\
Blackbody radiation shift               & $\m\0165.66^{\rm a}$     & 0.80    & $\m165.21^{\rm a}$     & 0.63\\
Relativistic redshift\\
and relativistic Doppler effect         & $-\0\085.56$     & $0.02^{\rm b}$    & $-\085.45$     & $0.02^{\rm b}$\\
Collisional shift                       & $-\0\0\06.1^{\rm a}$     & $2.4^{\rm a}$     & $\m\073.1^{\rm a}$     & $0.4^{\rm a}$\\
Distributed cavity phase shift          & $-\0\0\00.04$    & 0.93    & $-\0\00.28$    & 1.52\\
Microwave lensing												& $-\0\0\00.44$    & 0.20    & $-\0\00.67$    & 0.20\\
AC Stark shift (light shift)            & $\m\0\0\00.0$    & 0.01    & $\m\0\00.0$    & 0.01\\
Rabi and Ramsey pulling			& $\m\0\0\00.0$    & 0.013   & $\m\0\00.0$    & 0.013\\
Microwave leakage												& $\m\0\0\00.0$    & 0.01    & $\m\0\00.0$    & 0.01\\
Electronics											        & $\m\0\0\00.0$    & 0.1    & $\m\0\00.0$    & 0.1\\
Background gas pressure                 & $\m\0\0\00.0$    & 0.4     & $\m\0\00.0$    & 0.1\\
\mr
total                                   & $-1005.68$       & 2.74    & $-846.71$      & 1.71\\
\br
\end{tabular}
\item[] $^{\rm a}$ Typical numbers, which vary slightly for individual measurements.
\item[] $^{\rm b}$ For TAI contributions the uncertainty is $0.3\times10^{-16}$ (see subsection \ref{subsec:RRS}), which results in a slightly higher total uncertainty of $2.75\times10^{-16}$ for CSF1 and $1.74\times10^{-16}$ for CSF2.
\end{indented}
\end{table}

\section{Frequency instability}
\label{sec:Instabil}

\begin{figure}[t]
  \includegraphics[width=\columnwidth]{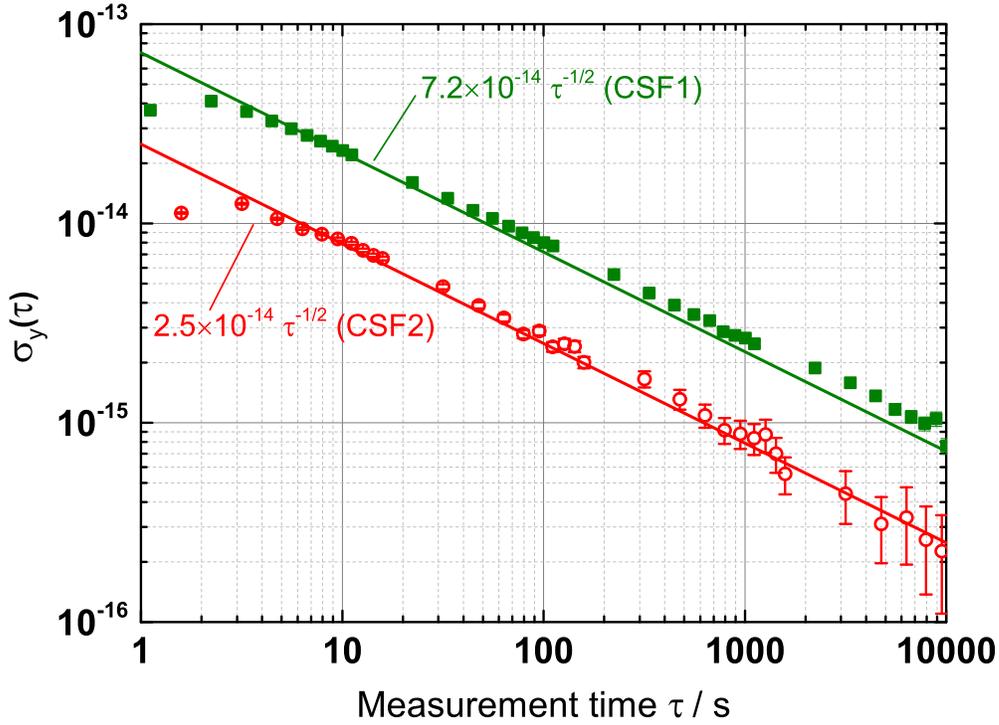}
	\caption{Allan standard deviation $\sigma_y(\tau)$ for high density operation of CSF1 and CSF2. The straight lines indicate the quantum projection noise after removing the noise contributions of the frequency references, a hydrogen maser for CSF1 and a $^{171}$Yb$^+$ single-ion frequency standard for CSF2. For averaging times $\tau<100$\,s, the reference noise contribution is negligible, since the OSMO is only slowly locked to the hydrogen maser and the $^{171}$Yb$^+$ frequency standard respectively \cite{Lipphardt2017}.}
	\label{fig:Allan} 
\end{figure}

Figure~\ref{fig:Allan} shows the quantum projection noise limited Allan standard deviation of CSF1 and CSF2 for high density operation. For CSF1, the reference signal contribution to the instability is small for averaging times $\tau$ less than $100$\,s, since the OSMO is slowly locked to the hydrogen maser with a time constant of $\approx50$\,s \cite{Lipphardt2017}. For longer averaging times the instability of the hydrogen maser leads to the deviation from $7.2 \times 10^{-14} (\tau/1\,\mathrm{s})^{-1/2}$. For CSF2, the OSMO was locked to the clock laser of an $^{171}$Yb$^+$ single-ion standard operating on the electric quadrupole transition \cite{Tamm2014}. Its noise contribution to the measured instability is barely visible in the graph at averaging times longer than $10$\,s. From figure~\ref{fig:Allan} we extract high-density frequency instabilities of $7.2 \times 10^{-14} (\tau/1\,\mathrm{s})^{-1/2}$ for CSF1 and $2.5 \times 10^{-14} (\tau/1\,\mathrm{s})^{-1/2}$ for CSF2. 

For normal operation, the overall frequency instabilities are degraded since the fountains are intermittently operated at low density (see subsection~\ref{subsec:Coll}). For CSF1, with a density ratio of about a factor of two, the overall frequency instability is typically $\sigma_y(\tau)=9.5 \times 10^{-14} (\tau/1\,\mathrm{s})^{-1/2}$. 

As mentioned in subsection~\ref{subsec:Coll}, the statistical uncertainty of the collisional shift evaluation of CSF2 significantly increases the effective frequency instability $\sigma_y^\mathrm{eff}(\tau)$. From $f_L$ and $f_H$ for low and high  density, the frequency $f_0 = 2f_L - f_H$ is corrected for the collisional shift. The statistical uncertainty $\delta f_0$ is $\delta f_0 = (4 \delta f_L^2 + \delta f_H^2)^{1/2}$, where $\delta f_L$ and $\delta f_H$ are the statistical uncertainties for low and high density. An overall measurement time $\tau$ gives $\delta f_L=\sigma_L\mathrm{(1\,s)}/(x\,\tau)^{1/2}$ and $\delta f_H=\sigma_H\mathrm{(1\,s)}/[(1-x)\,\tau]^{1/2}$ with $1$\,s Allan deviations $\sigma_{L(H)}\mathrm{(1\,s)}$, extrapolated to $\tau=1$\,s for low(high)-density operation, and $x$ is the fraction of time spent at low density and $1-x$ the time at high density. Accounting for the higher quantum projection noise for low density, $\sigma_L\mathrm{(1\,s)} = \sqrt{2}\,\sigma_H\mathrm{(1\,s)}$, $\delta f_0$ is minimized for $x\approx 2/3$. Because the frequency instabilities for low and high density are degraded by the maser instability, a typical effective frequency instability of CSF2 is $\sigma_y^\mathrm{eff}(\tau)=1.5 \times 10^{-13} (\tau/1\,\mathrm{s})^{-1/2}$, which is dominated by the statistical uncertainty of the collisional shift evaluation.

We note that it is possible to perform measurements, e.g. of an optical frequency for limited periods, $\tau \approx 1$\,d, operating only at high density. This provides the highest available frequency stability of CSF2, $2.5 \times 10^{-14} (\tau/1\,\mathrm{s})^{-1/2}$, which gives a statistical uncertainty $u_A\mathrm{(1\,d)}=9 \times 10^{-17}$. For such short periods, a collisional shift coefficient from previous and succeeding measurements can be applied, as in CSF1. The high stability of the measured shift coefficients of CSF2 (figure~\ref{fig:CollShift}) allow the collisional shift to be corrected with a systematic uncertainty at the low $10^{-16}$-level. Thus, the statistical uncertainty is less than the increased systematic uncertainty of $3\times 10^{-16}$, which is nonetheless almost a factor of two better than the overall 1-day uncertainty operating at high and low density as in subsection~\ref{subsec:Coll}.

\section{Applications of the PTB caesium fountains}
\label{sec:Applications}

Fountain clocks are typically utilized for calibrations of the TAI scale unit, steering of national time scales, and measurements of optical frequency standards. All of these applications benefit from the availability and reliability of fountain clocks. Moreover, they involve comparisons with other microwave and optical frequency standards and therefore provide independent information of the fountain performance and as tests of their accuracy evaluations. In the following we briefly compile the correspondent results of CSF1 and CSF2. 

\subsection{Calibrations of the TAI scale unit}
\label{sec:TAI} 

\begin{figure}[t]
  \includegraphics[width=\columnwidth]{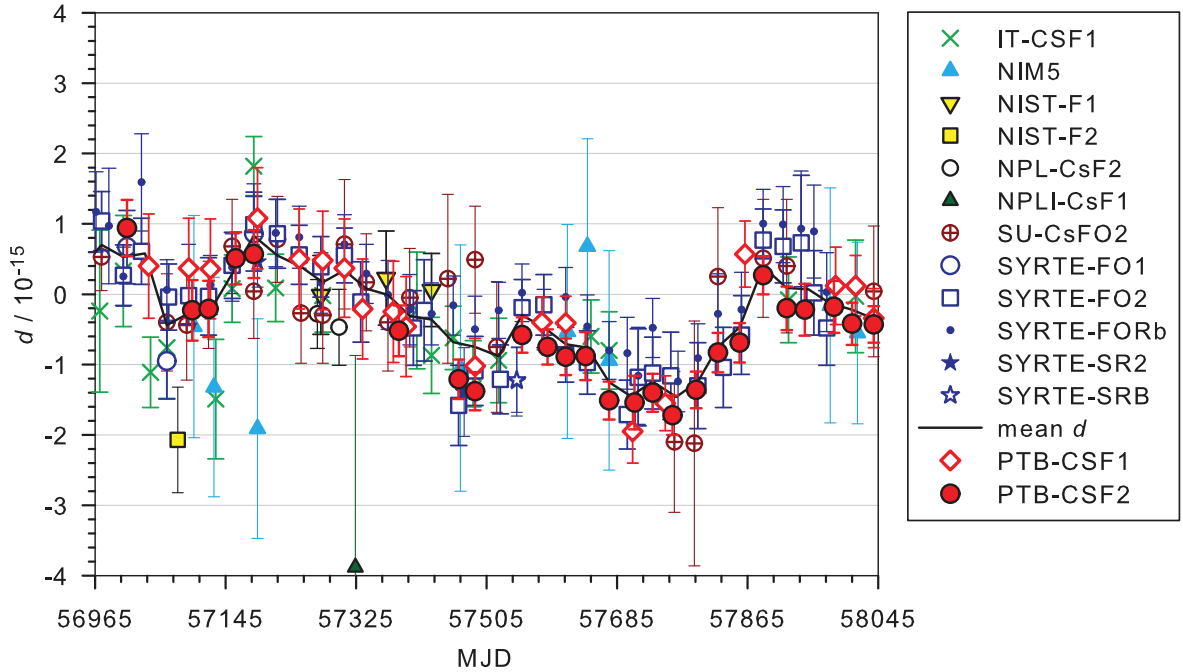}
	\caption{Fractional frequency difference $d$ between the scale unit of TAI and primary and secondary frequency standards, from three years of monthly calibration reports, November 2014 to October 2017 \cite{CircT}.  The measurement uncertainties are combined statistical and systematic uncertainties of the individual standards, local link uncertainties between the individual standards and clocks contributing to TAI, including  uncertainties due to dead-time, and individual link uncertainties to TAI. Mean $d$ is the estimate of $d$ by the BIPM based on all PFS and SFS measurements identified to be used for TAI steering over the respective period \cite{CircT}. MJD: Modified Julian Date.}
	\label{fig:TAI} 
\end{figure}

CSF1 and CSF2 have regularly contributed monthly calibrations of the TAI scale unit for many years\cite{CircT}. Recent calibrations, over three years from November 2014 through October 2017, are shown in figure~\ref{fig:TAI} for all contributing fountain clocks. Also shown are calibrations by secondary frequency standards (SFS), based on an optical Sr-transition frequency \cite{Lodewyck2016}. Both fountains have a significant weight in the steering of TAI due to their low statistical and systematic uncertainties, duty cycles of usually more than 90\%, a comparatively small link uncertainty to TAI, and their quite regular operation. Figure~\ref{fig:TAI} shows that CSF1 and CSF2 nicely agree with other standards and are close to the respective estimate of $d$, the monthly fractional frequency difference between the scale unit of TAI and primary and secondary frequency standards.

The mean frequency difference between CSF1 and CSF2 during twelve simultaneous TAI evaluations since 2016 is $2.2(1.5)\times 10^{-16}$. This result is compatible with the systematic uncertainties in Table~\ref{tab:uncertainties}, although the TAI evaluations of the two fountains were not fully congruent and not all of the investigations and techniques described here were completed or applied at the beginning of these comparisons.

\subsection{Steering of the time scale UTC(PTB)}
\label{sec:UTC} 

\begin{figure}[t]
  \includegraphics[width=\columnwidth]{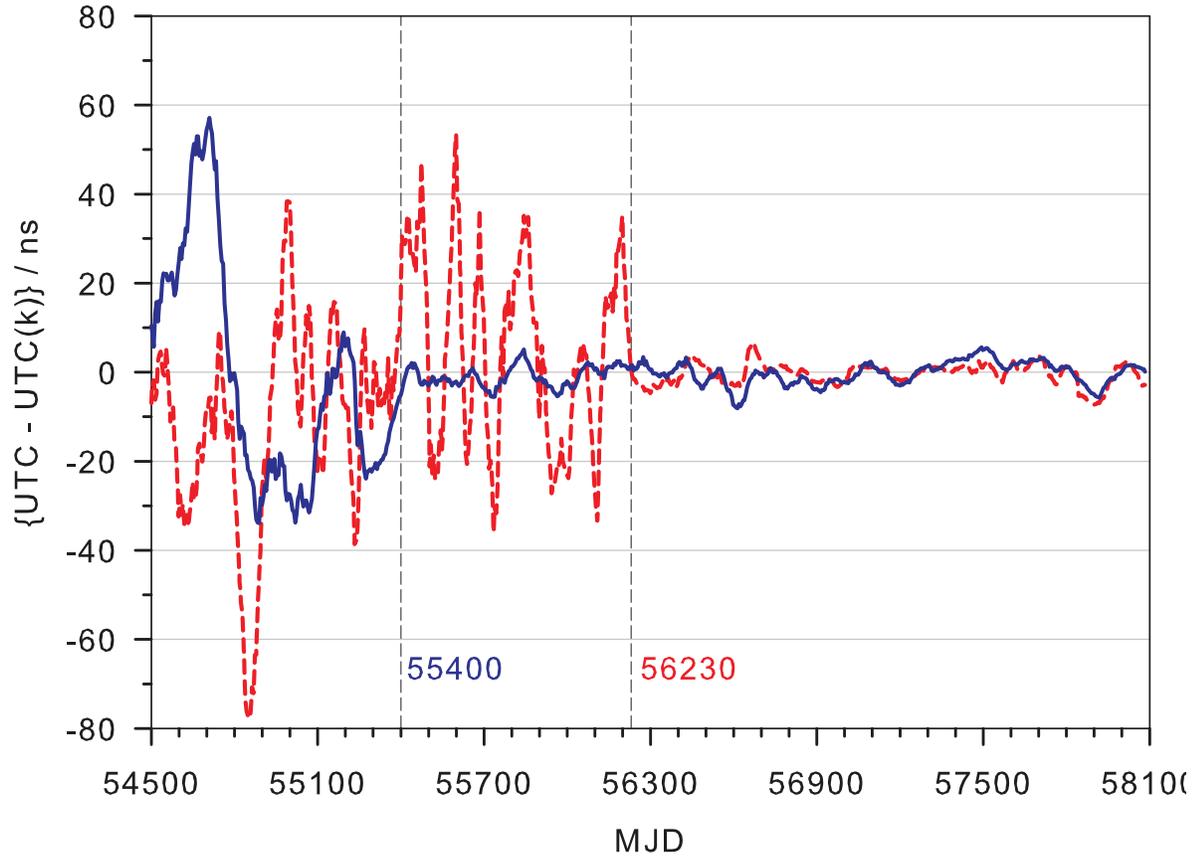}
	\caption{Time scale differences UTC-UTC(PTB) (solid blue curve) and UTC-UTC(OP) (dashed red curve), realized by LNE-SYRTE, over nearly 10 years, February 2008 through November 2017 \cite{CircT}.  The fountain data began steering UTC(PTB) around Modified Julian Date (MJD)\,55400 and around MJD\,56230 for UTC(OP).}
	\label{fig:UTC} 
\end{figure}

The time scale UTC(PTB) is the basis for legal time in Germany, Central European Time or Central European Summer Time. In the past UTC(PTB) was generated directly from the frequency output of the thermal-beam primary clock CS2, utilizing a phase micro stepper \cite{Bauch2004}. The deviations UTC-UTC(PTB) were usually below 50\,ns, but the day-to-day instability was inferior to some timescales based on hydrogen masers. Since 2010 UTC(PTB) uses a hydrogen maser that is steered daily by fountain measurements \cite{Bauch2012}. This new realization significantly improved the performance of UTC(PTB). While the day-to-day instability is now given by the hydrogen maser performance, figure~\ref{fig:UTC} shows that the deviations UTC-UTC(PTB) are now routinely below several nanoseconds \cite{CircT}. Since UTC is largely based on primary and secondary standards, the small deviations of UTC-UTC(PTB) also confirm the performance of CSF1 and CSF2.

\subsection{Optical frequency measurements}
\label{sec:OFM} 

Measurements of the frequencies of secondary optical frequency standards are needed to define their frequencies \cite{Gill2006}. Further, measurements of optical and microwave transition frequency ratios search for new physics, including temporal variations of fundamental constants, violations of Lorentz symmetry, and searches for light dark matter \cite{Huntemann2014,Hees2016,Safronova2018}. 

Several optical transition frequencies have been measured against the PTB fountain clocks. Two remote optical frequency measurements of the hydrogen 1S-2S and the $^{1}$S$_0$-$^{3}$P$_1$ transition frequency of $^{24}$Mg were performed via optical fibre links \cite{Matveev2013, Friebe2011}. Local measurements include the electric quadrupole (E2) and octupole (E3) optical clock transitions of a single-ion $^{171}$Yb$^+$ and the $^{1}$S$_0$-$^{3}$P$_0$ $^{87}$Sr optical lattice clock at PTB. These results have been reproducible and agree with measurements from other laboratories at a few parts in $10^{16}$ \cite{Tamm2014,Huntemann2012,Huntemann2014,Grebing2016}.

\section{Conclusions}
\label{sec:Con}
 
PTB's fountain clocks CSF1 and CSF2 have been steadily refined and improved. Here we report improved overall systematic uncertainties of $2.74 \times 10^{-16}$ for CSF1 and $1.71 \times 10^{-16}$ for CSF2 following a number of updated systematic uncertainty evaluations. Replacing the quartz-oscillator based microwave synthesis with one based on an optically stabilized microwave oscillator significantly improved the frequency stability, eliminating limitations from the Dick effect. Both fountains operate regularly and contribute to calibrations of TAI, the realization of the time scale UTC(PTB), and optical frequency measurements. These applications include direct and indirect comparisons with other primary and secondary standards and support the systematic evaluations of CSF1 and CSF2 reported here. Particularly noteworthy is the ascertained agreement at the low $10^{-16}$ level of CSF1, CSF2, and three fountain clocks of LNE-SYRTE in Paris, which were directly connected via an optical fibre link \cite{Guena2017}. Finally, we look forward to upcoming comparisons of the PTB fountain clocks CSF1 and CSF2 with other fountain and optical clocks within the Atomic Clocks Ensemble in Space (ACES) mission of the European Space Agency (ESA) \cite{Laurent2015}.

\ack

We acknowledge the contributions of many people at PTB to the development of CSF1 and CSF2 over the years. We wish to thank particularly R.~Schr\"oder, D.~Griebsch, U.~H\"ubner, A.~Bauch, C.~Tamm, R.~Wynands and N.~Nemitz for their significant and valuable contributions and N. Huntemann for providing the $^{171}$Yb$^+$ single-ion frequency standard reference signal. We are grateful for the constant technical support of C.~Richter, T.~Leder, M.~Menzel and A.~Hoppmann and valuable conversations with E.~Peik. We acknowledge a collaboration with H.~Denker, L.~Timmen and C.~Voigt from the Leibniz Universit\"at Hannover to re-evaluate the gravity potential at PTB and financial support from the European Metrology Research Programme (EMRP) in project SIB55~ITOC, the European Metrology Programme for Innovation and Research (EMPIR) in project 15SIB05~OFTEN, and the National Science Foundation. The EMRP and EMPIR are jointly funded by the EMRP/EMPIR participating countries within EURAMET and the European Union. 

\clearpage

\appendix

\section{Calculation of the Microwave Lensing Frequency Shift} \label{sec:AppMWL}

Many fountain clocks have two restrictive apertures, a lower state selection cavity aperture on the upward passage, and a single restrictive aperture on the downward passage, which clips the atoms before they are detected. Depending on the fountain design, the downward aperture is either the lower Ramsey cavity aperture or the state selection cavity aperture, if the detection zone is below the state selection cavity. CSF2, and other fountains \cite{Park2014}, have more than two restrictive apertures that clip the detected atoms, and each contributes to the microwave lensing frequency shift. Recent work on PHARAO, the laser-cooled caesium clock for the Atomic Clock Ensemble in Space (ACES) mission \cite{Laurent2015}, treated the microwave lensing shift from multiple apertures for a rectangular Ramsey cavity \cite{Peterman2016}. Here, CSF2 (and CSF1) have instead cylindrical Ramsey cavities and we similarly derive their associated microwave lensing frequency shifts.

The Rabi tipping angle, $\theta \left(\vec r \right)= \int {H_0 \left(\vec r \right) \mbox{d}z}$, is $\theta \left(r \right)= \theta \left(0 \right) J_0 \left(k r \right)$ for an azimuthally symmetric Ramsey cavity field \cite{Li2010}. We define the tipping angle in the first Ramsey interaction at $t_{1}$ as $\theta \left(r_1 \right)= \left(\pi/2\right) b_1 \eta  J_0 \left(k r_1 \right)$, where $b_1$ is an amplitude factor and $\eta =1.120$ \cite{Li2010}. In this way, a uniform atomic density illuminating the apertures yields a maximum Ramsey fringe contrast at $b_1= 1$, approximately a $\pi/2$-pulse, and $b_2$ similarly describes $\theta \left(r_2 \right)$ for the second Ramsey interaction at $t_{2}$. The velocity changes of the dressed states $\ket{1(2)}$ from the resonant dipole forces during the first Ramsey interaction are $\pm \delta \vec v\left({\vec r_{1}}\right) = b_1\eta\pi^2 \left({\nu_R/k} \right) J_1 \left({k r_{1}}\right) \hat{r_1}$ \cite{Li2011,Gibble2006,Gibble2014}, where the recoil shift is $\nu_R= h\nu_0^2/2mc^2$. Following \cite{Li2011,Gibble2006,Gibble2014,Peterman2016}, we semiclassically treat the atomic wave packets with ballistic trajectories, which have small deflections, of order nm/s, due to the microwave dipole forces from the first Ramsey interaction. The total difference of the detected dressed state populations $n_{1,2}$ gives the perturbation of the transition probability $\delta P$. Instead of integrating over the velocity distribution and the initial position distribution, it is more insightful to change variables to integrate over the apertures that are the most restrictive. We thus choose to first integrate over the two restrictive circular apertures, the bottom of the selection cavity waveguide aperture at time $t_{1L}$ and the bottom of either the Ramsey or state selection cavity waveguide aperture at $t_{2L}$ \cite{Gibble2014,Peterman2016}. Using $\vec r_{2L0}$ as the transverse atom position at $t_{2L}$, with no microwave lensing deflections at $t_{1}$, we get \cite{Li2011,Gibble2014,Peterman2016}:

\begin{eqnarray} 
\label{eq:Lensing1} 
\fl\delta P & = & -\frac{1}{4}\sum\limits_ \pm {\pm \int_{allspace}\!\!\!\!{n_0 \left({\vec r_{1L}, \vec r_{2L0}}\right)\sin \left[ {\theta \left({r_{2 \pm}} \right)} \right]\Theta \left( {a_2-r_{2L \pm}} \right)}}\Theta \left( {a_d-\left| {x_{d \pm}} \right|} \right)   \nonumber\\ 
\fl&& \times W_d\left({\vec r_{d \pm}} \right) \mbox{d}^2 r_{1L}\mbox{d}^2 r_{2L0}  
\nonumber\\
\fl&&\\
\fl\Delta P_R & = & \int_{allspace}\!\!\!\!{n_0\left({\vec r_{1L},\vec r_{2L0}}\right) \sin \left[ {\theta \left({r_1}\right)} \right]\sin \left[{\theta \left({r_{20}}\right)}\right]\Theta \left({a_2-r_{2L0}} \right)}\Theta \left({a_d-\left|{x_{d0}}\right|}\right) \nonumber\\ 
\fl&&\times W_d\left({\vec r_{d0}} \right) \mbox{d}^2 r_{1L} \mbox{d}^2 r_{2L0} \;\mbox{,} \nonumber\\
\fl&&\nonumber\\
\fl\mbox{where}&&\nonumber\\
\fl&&\nonumber\\	
\fl\vec{r}_{\beta\pm} & = & \vec{r}_{\beta 0}\pm \delta \vec{v}\left(r_1\right)\left(t_\beta-t_1 \right) \qquad \beta  \in \left\{2,2L,d \right\} \mbox{.} \nonumber
\end{eqnarray} 

\noindent Here we have explicitly written the microwave lensing deflections $\pm \delta \vec v\left({\vec r_{1}} \right)$, a ``$0$'' subscript denotes no lensing deflection, and we include the detection probability $W_d \left({\vec r_{d}} \right)$ \cite{Weyers2012,Li2011,Gibble2006,Gibble2014,Peterman2016}. The Ramsey fringe amplitude is $\Delta P_R$ and the microwave lensing frequency shift is $\delta \nu = \delta P/ \pi \left({t_2-t_1}\right)\Delta P_R$. The microwave lensing makes dressed state $\ket{1(2)}$ a little wider(narrower), so the Heaviside functions $\Theta \left({a_2-r_{2L\pm}} \right)$ and $\Theta \left({a_d -\left| {x_{d\pm}} \right|} \right)$, representing the apertures clipping the density $ n_0 \left( {\vec r_{1L} ,\vec r_{2L0} } \right)$, are effectively slightly narrower(wider).

We expand $\delta P$ in (\ref{eq:Lensing1}) to first order in the small velocity changes $\delta v\left({\vec r_{1}} \right)$ due to microwave lensing \cite{Weyers2012,Li2011,Gibble2006,Gibble2014,Peterman2016}. The cavity aperture at $t_{2L}$ and the detection masks at $t_d$ lead to three delta functions. It is simpler to integrate over the path of each aperture and therefore we use line integrals around the aperture at $t_{2L}$, integrating over $\phi_{2L0}$, and over the detection masks, $I_{det}$, using rectangular coordinates $\left(x_{d0}, y_{2L0}\right)$. We get the sum of three line integrals over these apertures, plus a surface integral \cite{Peterman2016}:

\begin{eqnarray} 
\label{eq:Lensing2} 
\fl\delta P&=& \frac{{a_2\left({t_{2L}-t_1}\right)}}{2}\int_{allspace}\!\!\!\!  \delta v\left({r_1} \right) n_0 \left( {\vec r_{1L} ,\vec r_{2L0}}\right)\sin \left[{\theta \left({r_{20}} \right)} \right]\Theta \left({a_d-\left| {x_{d0}} \right|} \right) \nonumber\\
\fl&&\times\left. {W_d \left({\vec r_{d0}} \right)\frac{{r_{2L0} \left({t_1-t_{1L}}\right) + r_{1L}\left({t_{2L}-t_1} \right)\cos \left({\phi_{2L0}} \right)}}{{r_1 \left({t_{2L}-t_{1L}}\right)}}} \right|_{r_{2L0}=a_2}\!\!\!\!\!\!\!\!\!\!\!\!\!\!\mbox{d}^2 r_{1L} \mbox{d}\phi _{2L0} \nonumber\\
\fl&&{}+I_{\det }+ \frac{{\nu_R}}{2}\int_{allspace}\!\!\!\!{n_0 \left({\vec r_{1L},\vec r_{2L0}} \right) \left. {\frac{{\partial \sin \left[{\theta \left({r_{2 -}}\right)} \right]W_d \left( {\vec r_{d-}} \right)}}{{\partial \nu_R}}}\right|_{\nu_R=0}} \nonumber \\
\fl&& \times \Theta \left(a_2-r_{2L0} \right)\Theta \left({a_d-\left|{x_{d0}}\right|} \right) \mbox{d}^2 r_{1L} \mbox{d}^2 r_{2L0}\\
\fl&&\nonumber\\	
\fl I_{\det}&=& \frac{{t_d-t_1}}{2}\frac{{t_{2L}-t_{1L} }}{{t_d-t_{1L}}} \sum \limits_\pm  {\pm \int_{allspace} \!\!\!\!\delta v_x \left( {\vec r_1} \right)n_0 \left({\vec r_{1L},\vec r_{2L0}} \right)\sin \left[{\theta \left({r_{20}}\right)}\right]} \nonumber \\ 
\fl&& \times\left. {\Theta \left({a_2-r_{2L0}} \right)W_d \left({\vec r_{d0}} \right)} \right|_{x_{2L0}=\frac{{\pm a_d\left({t_{2L}-t_{1L}} \right) +r_{1L}\cos\left({ \phi_{1L}}\right)\left({t_d-t_{2L}}\right)}} {{t_d-t_{1L}}}}\mbox{d}^2 r_{1L}\mbox{d}y_{2L0}\mbox{.}\nonumber
\end{eqnarray}

\noindent Here the sum over $\pm$ corresponds to the detection apertures at $\pm a_d$, and $\delta v_x=\delta v \left({r_1}\right) \cos \left({\phi_1} \right)$, with $\cos \left({\phi_{1}} \right)=x_1/r_1$.

Equations (\ref{eq:Lensing2}) can be significantly simplified with quite accurate approximations, which provides valuable checks. We can neglect the small detection variations and, near optimum amplitude, $b_{1,2}=1$, $\sin \left[ {\theta \left( {r_{1,2}}\right)}\right]$ has little variation so we can use $\sin \left( {b_{1,2}\,\eta\,\pi/2}\right)$ \cite{Li2011, Gibble2006,Gibble2014,Peterman2016}. With these approximations, the last term of (\ref{eq:Lensing2}), the surface integral, vanishes. In general, there are contributions from the detection apertures at $t_d$, in addition to those from the lower selection cavity aperture at $t_{1L}$ and the lower Ramsey or selection cavity aperture at $t_{2L}$ as for most fountain clocks \cite{Peterman2016}. This gives (\ref{eq:SimpLens}) in subsection~\ref{subsec:MWL}.

\newpage
\section*{References}
\bibliographystyle{iopart-num}
\bibliography{CSF2017}

\providecommand{\newblock}{}
\begin{thebibliography}{10}
\expandafter\ifx\csname url\endcsname\relax
  \def\url#1{{\tt #1}}\fi
\expandafter\ifx\csname urlprefix\endcsname\relax\def\urlprefix{URL }\fi
\providecommand{\eprint}[2][]{\url{#2}}

\bibitem{Wynands2005}
Wynands R and Weyers S 2005 {\em Metrologia\/} {\bf 42} S64
  \urlprefix\url{http://stacks.iop.org/0026-1394/42/i=3/a=S08}

\bibitem{CircT}
{BIPM Circular T} \urlprefix\url{http://www.bipm.org/jsp/en/TimeFtp.jsp}

\bibitem{Weyers2001}
Weyers S, H\"ubner U, Schr\"oder R, Tamm C and Bauch A 2001 {\em Metrologia\/}
  {\bf 38} 343 \urlprefix\url{http://stacks.iop.org/0026-1394/38/i=4/a=7}

\bibitem{Gerginov2010}
Gerginov V, Nemitz N, Weyers S, Schr\"oder R, Griebsch D and Wynands R 2010
  {\em Metrologia\/} {\bf 47} 65
  \urlprefix\url{http://stacks.iop.org/0026-1394/47/i=1/a=008}

\bibitem{Szymaniec2010}
Szymaniec K, Park S~E, Marra G and Chalupczak W 2010 {\em Metrologia\/} {\bf
  47} 363 \urlprefix\url{http://stacks.iop.org/0026-1394/47/i=4/a=003}

\bibitem{Guena2012}
Gu{\'e}na J, Abgrall M, Rovera D, Laurent P, Chupin B, Lours M, Santarelli G,
  Rosenbusch P, Tobar M~E, Li R, Gibble K, Clairon A and Bize S 2012 {\em IEEE
  Trans. Ultrason., Ferroelectr., Freq. Control\/} {\bf 59} 391--410 ISSN
  0885-3010

\bibitem{Domnin2013}
Domnin Y~S, Baryshev V~N, Boyko A~I, Elkin G~A, Novoselov A~V, Kopylov L~N and
  Kupalov D~S 2013 {\em Meas. Tech.\/} {\bf 55} 1155--1162 ISSN 1573-8906
  \urlprefix\url{http://dx.doi.org/10.1007/s11018-012-0102-0}

\bibitem{Levi2014}
Levi F, Calonico D, Calosso C~E, Godone A, Micalizio S and Costanzo G~A 2014
  {\em Metrologia\/} {\bf 51} 270
  \urlprefix\url{http://stacks.iop.org/0026-1394/51/i=3/a=270}

\bibitem{Jefferts2002}
Jefferts S~R, Shirley J, Parker T~E, Heavner T~P, Meekhof D~M, Nelson C, Levi
  F, Costanzo G, Marchi A~D, Drullinger R, Hollberg L, Lee W~D and Walls F~L
  2002 {\em Metrologia\/} {\bf 39} 321
  \urlprefix\url{http://stacks.iop.org/0026-1394/39/i=4/a=1}

\bibitem{Heavner2014}
Heavner T~P, Donley E~A, Levi F, Costanzo G, Parker T~E, Shirley J~H, Ashby N,
  Barlow S and Jefferts S~R 2014 {\em Metrologia\/} {\bf 51} 174
  \urlprefix\url{http://stacks.iop.org/0026-1394/51/i=3/a=174}

\bibitem{Gibble2015}
Gibble K 2015 {\em Metrologia\/} {\bf 52} 163
  \urlprefix\url{http://stacks.iop.org/0026-1394/52/i=1/a=163}

\bibitem{Fang2015}
Fang F, Li M, Lin P, Chen W, Liu N, Lin Y, Wang P, Liu K, Suo R and Li T 2015
  {\em Metrologia\/} {\bf 52} 454
  \urlprefix\url{http://stacks.iop.org/0026-1394/52/i=4/a=454}

\bibitem{Acharya2017}
Acharya A, Bharath V, Arora P, Yadav S, Agarwal A and Gupta A~S 2017 {\em
  MAPAN-Journal of Metrology Society of India\/} {\bf 32} 67 -- 76

\bibitem{Guena2014}
Gu\'ena J, Abgrall M, Clairon A and Bize S 2014 {\em Metrologia\/} {\bf 51} 108
  \urlprefix\url{http://stacks.iop.org/0026-1394/51/i=1/a=108}

\bibitem{Huntemann2014}
Huntemann N, Lipphardt B, Tamm C, Gerginov V, Weyers S and Peik E 2014 {\em
  Phys. Rev. Lett.\/} {\bf 113}(21) 210802
  \urlprefix\url{http://link.aps.org/doi/10.1103/PhysRevLett.113.210802}

\bibitem{Tamm2014}
Tamm C, Huntemann N, Lipphardt B, Gerginov V, Nemitz N, Kazda M, Weyers S and
  Peik E 2014 {\em Phys. Rev. A\/} {\bf 89}(2) 023820
  \urlprefix\url{http://link.aps.org/doi/10.1103/PhysRevA.89.023820}

\bibitem{Grebing2016}
Grebing C, Al-Masoudi A, D\"{o}rscher S, H\"{a}fner S, Gerginov V, Weyers S,
  Lipphardt B, Riehle F, Sterr U and Lisdat C 2016 {\em Optica\/} {\bf 3}
  563--569
  \urlprefix\url{http://www.osapublishing.org/optica/abstract.cfm?URI=optica-3-6-563}

\bibitem{Matveev2013}
Matveev A, Parthey C~G, Predehl K, Alnis J, Beyer A, Holzwarth R, Udem T,
  Wilken T, Kolachevsky N, Abgrall M, Rovera D, Salomon C, Laurent P, Grosche
  G, Terra O, Legero T, Schnatz H, Weyers S, Altschul B and H\"ansch T~W 2013
  {\em Phys. Rev. Lett.\/} {\bf 110}(23) 230801
  \urlprefix\url{https://link.aps.org/doi/10.1103/PhysRevLett.110.230801}

\bibitem{Friebe2011}
Friebe J, Riedmann M, W\"ubbena T, Pape A, Kelkar H, Ertmer W, Terra O, Sterr
  U, Weyers S, Grosche G, Schnatz H and Rasel E~M 2011 {\em New J. Phys.\/}
  {\bf 13} 125010
  \urlprefix\url{http://stacks.iop.org/1367-2630/13/i=12/a=125010}

\bibitem{Bauch2012}
Bauch A, Weyers S, Piester D, Staliuniene E and Yang W 2012 {\em Metrologia\/}
  {\bf 49} 180 \urlprefix\url{http://stacks.iop.org/0026-1394/49/i=3/a=180}

\bibitem{Guena2017}
Gu\'ena J, Weyers S, Abgrall M, Grebing C, Gerginov V, Rosenbusch P, Bize S,
  Lipphardt B, Denker H, Quintin N, Raupach S~M~F, Nicolodi D, Stefani F,
  Chiodo N, Koke S, Kuhl A, Wiotte F, Meynadier F, Camisard E, Chardonnet C,
  {Le Coq} Y, Lours M, Santarelli G, Amy-Klein A, {Le Targat} R, Lopez O,
  Pottie P~E and Grosche G 2017 {\em Metrologia\/} {\bf 54} 348
  \urlprefix\url{http://stacks.iop.org/0026-1394/54/i=3/a=348}

\bibitem{Weyers2001b}
Weyers S, Bauch A, Schr\"oder R and Tamm C 2001 {The atomic caesium fountain
  CSF1 of PTB} {\em {Proceedings of the 6th Symposium on Frequency Standards
  and Metrology}\/} ed Gill P (University of St Andrews, Fife, Scotland) pp
  64--71

\bibitem{Schroeder2002}
Schr\"oder R, H\"ubner U and Griebsch D 2002 {\em IEEE Trans. Ultrason.,
  Ferroelectr., Freq. Control\/} {\bf 49} 383--392 ISSN 0885-3010

\bibitem{Lu1996}
Lu Z~T, Corwin K~L, Renn M~J, Anderson M~H, Cornell E~A and Wieman C~E 1996
  {\em Phys. Rev. Lett.\/} {\bf 77} 3331--3334
  \urlprefix\url{https://link.aps.org/doi/10.1103/PhysRevLett.77.3331}

\bibitem{Dobrev2016}
Dobrev G, Gerginov V and Weyers S 2016 {\em Phys. Rev. A\/} {\bf 93}(4) 043423
  \urlprefix\url{http://link.aps.org/doi/10.1103/PhysRevA.93.043423}

\bibitem{PereiraDosSantos2002}
Pereira Dos~Santos F, Marion H, Bize S, Sortais Y, Clairon A and Salomon C 2002
  {\em Phys. Rev. Lett.\/} {\bf 89} 233004
  \urlprefix\url{https://link.aps.org/doi/10.1103/PhysRevLett.89.233004}

\bibitem{Kazda2013}
Kazda M, Gerginov V, Nemitz N and Weyers S 2013 {\em IEEE Trans. Instrum.
  Meas.\/} {\bf 62} 2812--2819 ISSN 0018-9456

\bibitem{Lipphardt2017}
Lipphardt B, Gerginov V and Weyers S 2017 {\em IEEE Trans. Ultrason.,
  Ferroelectr., Freq. Control\/} {\bf 64} 761--766 ISSN 0885-3010

\bibitem{Gupta2000}
Sen~Gupta A, Popovic D and Walls F~L 2000 {\em IEEE Trans. Ultrason.,
  Ferroelectr., Freq. Control\/} {\bf 47} 475--479 ISSN 0885-3010

\bibitem{Gupta2007}
Sen~Gupta A, Schr\"oder R, Weyers S and Wynands R 2007 {A New 9-\textsc{GH}z
  Synthesis Chain for Atomic Fountain Clocks} {\em Proceedings of the 2007
  Joint Meeting of the European Frequency and Time Forum and the IEEE
  International Frequency Control Symposium\/} (Geneva, Switzerland) pp
  234--237

\bibitem{Kazda2018}
Kazda M 2018 {\em Advanced Microwave Control for Atomic Fountain Clocks\/}
  Ph.D. thesis Technische Universit\"at Braunschweig (Germany)

\bibitem{Schroeder1991}
Schr\"oder R 1991 {Frequency Synthesis in Primary Cesium Clocks} {\em
  Proceedings of the 5th European Frequency and Time Forum\/} (Noordwijk, The
  Netherlands) pp 194--200

\bibitem{Santarelli1998}
Santarelli G, Audoin C, Makdissi A, Laurent P, Dick G and Clairon A 1998 {\em
  IEEE Trans. Ultrason., Ferroelectr., Freq. Control\/} {\bf 45} 887--894

\bibitem{Weyers2012}
Weyers S, Gerginov V, Nemitz N, Li R and Gibble K 2012 {\em Metrologia\/} {\bf
  49} 82 \urlprefix\url{http://stacks.iop.org/0026-1394/49/i=1/a=012}

\bibitem{Weyers2000}
Weyers S, Bauch A, H\"ubner U, Schr\"oder R and Tamm C 2000 {\em IEEE Trans.
  Ultrason., Ferroelectr., Freq. Control\/} {\bf 47} 432 -- 437

\bibitem{Rosenbusch2007}
Rosenbusch P, Zhang S and Clairon A {2007} {Blackbody radiation shift in
  primary frequency standards} {\em {Proceedings of the 2007 Joint Meeting of
  the European Frequency and Time Forum and the IEEE International Frequency
  Control Symposium}\/} (Geneva, Switzerland) pp {1060--1063} {}

\bibitem{Angstmann2006}
Angstmann E~J, Dzuba V~A and Flambaum V~V 2006 {\em Phys. Rev. A\/} {\bf 74}(2)
  023405 \urlprefix\url{http://link.aps.org/doi/10.1103/PhysRevA.74.023405}

\bibitem{Denker2018}
Denker H, Timmen L, Voigt C, Weyers S, Peik E, Margolis H~S, Delva P, Wolf P
  and Petit G 2018 {\em Journal of Geodesy\/} {\bf 92} 487--516 ISSN 1432-1394
  \urlprefix\url{https://doi.org/10.1007/s00190-017-1075-1}

\bibitem{Tiesinga1992}
Tiesinga E, Verhaar B~J, Stoof H~T~C and van Bragt D 1992 {\em Phys. Rev. A\/}
  {\bf 45} R2671--R2673
  \urlprefix\url{https://link.aps.org/doi/10.1103/PhysRevA.45.R2671}

\bibitem{Gibble1993}
Gibble K and Chu S 1993 {\em Phys. Rev. Lett.\/} {\bf 70} 1771--1774

\bibitem{Szymaniec2007a}
Szymaniec K, Chalupczak W, Tiesinga E, Williams C~J, Weyers S and Wynands R
  2007 {\em Phys. Rev. Lett.\/} {\bf 98} 153002
  \urlprefix\url{https://link.aps.org/doi/10.1103/PhysRevLett.98.153002}

\bibitem{Marion2004}
Marion H, Bize S, Cacciapuoti L, Chambon D, Pereira~dos Santos F, Santarelli G,
  Wolf P, Clairon A, Luiten A, Tobar M, Kokkelmans S and Salomon C 2004 First
  observation of {Feshbach} resonances at very low magnetic field in a
  $^{133}${Cs} fountain {\em Proceedings of the 18th European Frequency and
  Time Forum\/} (Guildford, UK) pp 49--55

\bibitem{Papoular2012}
Papoular D~J, Bize S, Clairon A, Marion H, Kokkelmans S~J~J~M~F and Shlyapnikov
  G~V 2012 {\em Phys. Rev. A\/} {\bf 86}(4) 040701
  \urlprefix\url{https://link.aps.org/doi/10.1103/PhysRevA.86.040701}

\bibitem{Bennett2017}
See~Bennett A, Gibble K, Kokkelmans S and Hutson J~M 2017 {\em Phys. Rev.
  Lett.\/} {\bf 119}(11) 113401, and references therein
  \urlprefix\url{https://link.aps.org/doi/10.1103/PhysRevLett.119.113401}

\bibitem{Li2004}
Li R and Gibble K 2004 {\em Metrologia\/} {\bf 41} 376

\bibitem{Li2010}
Li R and Gibble K 2010 {\em Metrologia\/} {\bf 47} 534

\bibitem{Guena2011}
Gu\'ena J, Li R, Gibble K, Bize S and Clairon A 2011 {\em Phys. Rev. Lett.\/}
  {\bf 106}(13) 130801
  \urlprefix\url{https://link.aps.org/doi/10.1103/PhysRevLett.106.130801}

\bibitem{Li2011}
Li R, Gibble K and Szymaniec K 2011 {\em Metrologia\/} {\bf 48} 283
  \urlprefix\url{http://stacks.iop.org/0026-1394/48/i=5/a=007}

\bibitem{Nemitz2012}
Nemitz N, Gerginov V, Wynands R and Weyers S 2012 {\em Metrologia\/} {\bf 49}
  468 \urlprefix\url{http://stacks.iop.org/0026-1394/49/i=4/a=468}

\bibitem{Li2005}
Li R and Gibble K 2005 {Distributed cavity phase and the associated power
  dependence} {\em Proceedings of the Joint IEEE International Frequency
  Control Symposium and Precise Time and Time Interval Systems and Applications
  Meeting\/} (Vancouver, BC, Canada) pp 99--104

\bibitem{Weyers2007}
Weyers S, Wynands R, Szymaniec K and Chalupczak W 2007 {Multiple $\pi/2$ pulse
  area operation of caesium fountains and the collisional frequency shift} {\em
  Proceedings of the IEEE International Frequency Control Symposium Joint with
  the 21st European Frequency and Time Forum\/} (Geneva, Switzerland) pp 52--54

\bibitem{Gerginov2010a}
Gerginov V, Nemitz N, Griebsch D, Kazda M, Li R, Gibble K, Wynands R and Weyers
  S 2010 {Recent improvements and current uncertainty budget of PTB fountain
  clock CSF2} {\em Proceedings of the 24th European Frequency and Time Forum\/}
  (Noordwijk, The Netherlands)

\bibitem{Bloom1962}
Bloom M and Erdman K 1962 {\em Can. J. Phys.\/} {\bf 40} 179--193
  \urlprefix\url{https://doi.org/10.1139/p62-016}

\bibitem{Sleator1992}
Sleator T, Pfau T, Balykin V, Carnal O and Mlynek J 1992 {\em Phys. Rev.
  Lett.\/} {\bf 68}(13) 1996--1999
  \urlprefix\url{https://link.aps.org/doi/10.1103/PhysRevLett.68.1996}

\bibitem{Gibble2006}
Gibble K 2006 {\em Phys. Rev. Lett.\/} {\bf 97} 073002

\bibitem{Gibble2016}
Gibble K 2016 {Systematic Effects in Atomic Fountain Clocks} {\em {Proceedings
  of the 8th Symposium on Frequency Standards and Metrology 2015}\/} vol 723 ed
  Riehle F (Potsdam, Germany) p 012002

\bibitem{Gibble2014}
Gibble K 2014 {\em Phys. Rev. A\/} {\bf 90}(1) 015601

\bibitem{Peterman2016}
Peterman P, Gibble K, Laurent P and Salomon C 2016 {\em Metrologia\/} {\bf 53}
  899 \urlprefix\url{http://stacks.iop.org/0026-1394/53/i=2/a=899}

\bibitem{Vanier1989}
Vanier J and Audoin C 1989 {\em The Quantum Physics of Atomic Frequency
  Standards\/} (Adam Hilger, Bristol and Philadelphia)

\bibitem{Gerginov2014}
Gerginov V, Nemitz N and Weyers S 2014 {\em Phys. Rev. A\/} {\bf 90}(3) 033829
  \urlprefix\url{http://link.aps.org/doi/10.1103/PhysRevA.90.033829}

\bibitem{Majorana1932}
Majorana E 1932 {\em Il Nuovo Cimento\/} {\bf 9} 43--50

\bibitem{Bauch1993}
Bauch A and Schr\"oder R 1993 {\em Annalen der Physik\/} {\bf 505} 421--449
  \urlprefix\url{http://dx.doi.org/10.1002/andp.19935050502}

\bibitem{Wynands2007}
Wynands R, Schr\"oder R and Weyers S 2007 {\em IEEE Trans. Instrum. Meas.\/}
  {\bf 56} 660 -- 663

\bibitem{Boussert1998}
Boussert B, Theobald G, Cerez P and de~Clercq E 1998 {\em IEEE Trans. Ultrason.
  Ferroelectr. Freq. Control\/} {\bf 45} 728--738 ISSN 0885-3010

\bibitem{Weyers2006}
Weyers S, Schr\"oder R and Wynands R 2006 Effects of microwave leakage in
  caesium clocks: Theoretical and experimental results {\em Proceedings of the
  20th European Frequency and Time Forum\/} pp 173--180

\bibitem{Shirley2006}
Shirley J~H, Levi F, Heavner T~P, Calonico D, Yu D~H and Jefferts S~R 2006 {\em
  IEEE Trans. Ultrason. Ferroelectr. Freq. Control\/} {\bf 53} 2376--2385 ISSN
  0885-3010

\bibitem{Santarelli2009}
Santarelli G, Governatori G, Chambon D, Lours M, Rosenbusch P, Guena J,
  Chapelet F, Bize S, Tobar M, Laurent P, Potier T and Clairon A 2009 {\em IEEE
  Trans. Ultrason., Ferroelectr., Freq. Control\/} {\bf 56} 1319 -- 1326

\bibitem{Lin2009}
Lin P, Liu S and Liu N 2009 Microwave leakage shift suppression based on home
  made {DDS} {\em Proceedings of the 2009 IEEE International Frequency Control
  Symposium Joint with the 22nd European Frequency and Time forum\/} pp
  1026--1029 ISSN 2327-1914

\bibitem{Kazda2016a}
Kazda M and Gerginov V 2016 {\em IEEE Trans. Instrum. Meas.\/} {\bf 65}
  2389--2393 ISSN 0018-9456

\bibitem{Kazda2018a}
Kazda M, Gerginov V, Lipphardt B and Weyers S 2018 {\em in preparation\/}

\bibitem{Ramsey1955}
Ramsey N~F 1955 {\em Phys. Rev.\/} {\bf 100}(4) 1191--1194

\bibitem{Shirley1963}
Shirley J~H 1963 {\em J. Appl. Phys.\/} {\bf 34} 783--788
  \urlprefix\url{http://dx.doi.org/10.1063/1.1729536}

\bibitem{Audoin1978}
Audoin C, Jardino M, Cutler L~S and Lacey R~F 1978 {\em IEEE Trans. Instrum.
  Meas.\/} {\bf {27}} {325--329} ISSN {0018-9456}

\bibitem{Levi2006a}
Levi F, Shirley J~H, Heavner T~P, Yu D~H and Jefferts S~R 2006 {\em IEEE Trans.
  Ultrason., Ferroelectr., Freq. Control\/} {\bf 53} 1584 -- 1589

\bibitem{Shirley2009}
Shirley J~H, Heavner T~P and Jefferts S~R 2009 {\em IEEE Trans. Instrum.
  Meas.\/} {\bf {58}} {1241--1246} ISSN {0018-9456, 1557-9662}

\bibitem{Heavner2006}
Heavner T~P, Shirley J~H, Levi F, Yu D and Jefferts S~R 2006 Frequency biases
  in pulsed atomic fountain frequency standards due to spurious components in
  the microwave spectrum {\em 2006 IEEE International Frequency Control
  Symposium and Exposition\/} pp 273--276 ISSN 2327-1914

\bibitem{Kazda2016}
Kazda M, Gerginov V, Huntemann N, Lipphardt B and Weyers S 2016 {\em IEEE
  Trans. Ultrason., Ferroelectr., Freq. Control\/} {\bf 63} 970--974 ISSN
  0885-3010

\bibitem{Gibble2013}
Gibble K 2013 {\em Phys. Rev. Lett.\/} {\bf 110}(18) 180802

\bibitem{Lodewyck2016}
Lodewyck J, Bilicki S, Bookjans E, Robyr J~L, Shi C, Vallet G, Targat R~L,
  Nicolodi D, Le~Coq Y, Guéna J, Abgrall M, Rosenbusch P and Bize S 2016 {\em
  Metrologia\/} {\bf 53} 1123
  \urlprefix\url{http://stacks.iop.org/0026-1394/53/i=4/a=1123}

\bibitem{Bauch2004}
Bauch A, Piester D and Staliuniene E 2004 A new realization strategy for the
  time scale {UTC(PTB)} {\em Proceedings of the 2004 IEEE International
  Frequency Control Symposium and Exposition, 2004.\/} pp 518--523 ISSN
  1075-6787

\bibitem{Gill2006}
Gill P and Riehle F 2006 On secondary representations of the second {\em
  Proceedings of the 20th European Frequency and Time Forum\/} (Braunschweig,
  Germany) pp 282--288

\bibitem{Hees2016}
Hees A, Gu\'ena J, Abgrall M, Bize S and Wolf P 2016 {\em Phys. Rev. Lett.\/}
  {\bf 117}(6) 061301
  \urlprefix\url{https://link.aps.org/doi/10.1103/PhysRevLett.117.061301}

\bibitem{Safronova2018}
Safronova M~S, Budker D, DeMille D, {Jackson Kimball} D~F, Derevianko A and
  Clark C~W 2018 {\em Rev. Mod. Phys.\/} {\bf 90}(2) 025008
  \urlprefix\url{https://link.aps.org/doi/10.1103/RevModPhys.90.025008}

\bibitem{Huntemann2012}
Huntemann N, Okhapkin M, Lipphardt B, Weyers S, Tamm C and Peik E 2012 {\em
  Phys. Rev. Lett.\/} {\bf 108}(9) 090801
  \urlprefix\url{https://link.aps.org/doi/10.1103/PhysRevLett.108.090801}

\bibitem{Laurent2015}
Laurent P, Massonnet D, Cacciapuoti L and Salomon C 2015 {\em C. R. Phys.\/}
  {\bf 16} 540--552 \urlprefix\url{https://doi.org/10.1016/j.crhy.2015.05.002}

\bibitem{Park2014}
Park S~E, Heo M~S, Kwon T~Y, Gibble K, Lee S~B, Park C~Y, Lee W~K and Yu D~H
  2014 {Accuracy evaluation of the KRISS-F1 fountain clock} {\em 2014 IEEE
  International Frequency Control Symposium (FCS)\/} pp 1--2 ISSN 2327-1914
  \urlprefix\url{http://dx.doi.org/10.1109/FCS.2014.6859971}

\end{thebibliography}

\end{document}